# Layer-by-layer connection for large area single crystal boron nitride multilayer films

Hui Shi[1#], Mingyuan Wang[1#], Hongying Chen[2#], Adrien Rousseau[3#], Junpeng Shu[1], Ming Tian[1], Ruowang Chen[1], Juliette Plo[3], Pierre Valvin[3], Bernard Gil[3], Jiajie Qi[4], Qinghe Wang[4], Kaihui Liu[4*], Mingliang Zhang[1], Guillaume Cassabois[3*], Di Wu[2*], Neng Wan[1*]

1, Key laboratory of MEMS of Ministry of Education, School of Integrated Circuits, Southeast University, Nanjing 210096, China

2, National Laboratory of Solid State Microstructures, Department of Materials Science and Engineering, Jiangsu Key Laboratory of Artificial Functional Materials, and Collaborative Innovation Center of Advanced Microstructures, Nanjing University, Nanjing 210093, China

3, Laboratoire Charles Coulomb, UMR5221 CNRS-Université de Montpellier, 34095 Montpellier, France

4, State Key Laboratory for Mesoscopic Physics, Frontiers Science Center for Nano-optoelectronics, School of Physics, Peking University, Beijing 100871, China

# Contributed equally

*E-mail: khliu@pku.edu.cn; guillaume.cassabois@umontpellier.fr; diwu@nju.edu.cn; wn@seu.edu.cn

# Abstract

Boron nitride (BN) is today considered as one of the most promising materials for many novel applications including bright single photon emission, deep UV opto-electronics, small sized solid-state neutron detector, and high-performance two-dimensional materials, etc. Despite the recent successful fabrication of large-area BN single-crystals (typically ≤ 5 atomic layers), the scalable growth of thicker single-crystalline BN films still constitutes a great challenge. In this work, we demonstrate an approach to grow large-area multilayer single-crystal BN films by chemical vapor deposition on face-centered cubic Fe-Ni (111) single crystal alloy thin films with different stoichiometric phases. We show that the BN growth is greatly tunable and improved by increasing the Fe content in single-crystal Fe-Ni (111). The formation of pyramid-shaped multilayer BN domains with aligned orientation enables a continuous connection following a layer-by-layer, "first-meet-first-connect", mosaic stitching mechanism. By means of selected area electron diffraction, micro-photoluminescence spectroscopy in the deep UV and high-resolution transmission electron microscopy, the layer-by-layer connection mechanism is unambiguously evidenced, and the stacking order has been verified to occur as unidirectional AB and ABC stackings, i.e., in the Bernal and rhombohedral BN phase.

# Introduction

Two-dimensional boron nitride (BN) materials have been attracting extensive attention in recent years due to their novel properties and potential applications in many fields including bright single photon emitters in the visible [1] and ultraviolet regions [2]; high efficient deep UV emission matrix [8]; small-sized solid state neutron detectors [9]; it also works as excellent two-dimensional material substrate [3, 4] and protective capping layer [4 - 7], , for achieving outstanding device performance [10-12]. One of the most important prerequisites for these applications is the scalable fabrication of high-quality single-crystal BN films. Recent studies have found ways to grow large-area single-crystal BN monolayers [13, 14] using chemical vapor deposition (CVD) methods. However, monolayer CVD-grown BN tends to be defective (thus leading to high current-leakage [15] and poor screening in electronic devices), and still easy to be ruptured or polluted during the mandatory transfer process for any practical use. This hinders the final application of the single layer CVD-grown BN. Alternatively, multilayers of single-crystal BN with a large-area uniformity are highly desirable for most applications in order to reduce leakage, enhance screening, and increase mechanical stability [5, 16 - 18], with a potential surface cleaning process by simply removing the first few polluted layers.

Reminding the long-reported near-perfect epitaxy relationship between BN and a single crystal nickel [19], also quite similar to the mechanism of monolayer BN growth on single-crystal copper [13, 14], BN single-crystals with a thickness ≤ 5 atomic layers were recently reported [20]. A mechanism with individual nucleation followed by mosaic stitching (INMS) is observed. The multilayers are shown as 2, 3, 5-layered BN with AA'-type stacking, indicating a hexagonal BN phase (hBN) as the lowest energy state. However, single-crystal multilayers thicker than 5 atomic layers (~ 2 nm film thickness) are reported to be still very hard to be synthesized even at relatively high growth temperatures. In reports by H. Ago group [4, 21, 22], BN multilayers with thickness of 2 - 10 nm were fabricated by using Fe-Ni alloys. The BN films showed good uniformity. However, the single crystallinity and the stacking structure were not evidenced.

The growth of multilayer BN based on the INMS mechanism faces several difficulties. First, successful implementation of the INMS mechanism requires that all isolated nucleated domains have the same orientation, as also demonstrated for the monolayer BN growth [13, 14]. Second, specifically for multilayer BN, the adjacent BN domains have to connect their layers correctly, in order to achieve a single-crystal multilayer (Figs. 1 (a), (b)). Otherwise, detrimental grain boundaries or broken edges form between the randomly encountered nucleated islands (Fig 1(a)). In order to fulfill these two conditions, the alignment, the shape and the stacking order of individual BN domains have to be well controlled.

We here demonstrate a strategy for achieving large-area multilayer BN single-crystal films. It is known that multilayer BN domains tend to grow following a "bottom nucleate" habitation/ mechanism [23], where the new BN layer (that beneath the as-grown layers) nucleates at the interface between the as-grown BN layer and the metal (Fig. 1 (c)) at the spire of the pyramids [23 - 25]. This growth habitation normally leads to two different structures according to previous reports, with different stacking order, e.g., either the AA' stacking, or the AB (or ABC) stacking, which finally appeared as two different structures as hexagram (AA', Fig. 1 (a)) [25] or pyramid (AB or ABC, Fig. 1 (b)) shapes, respectively [23 - 25].

It is then realized that a hexagram might not be a suitable shape for the formation of single-crystal BN multilayers following INMS due to the high possibility of forming anti-domain boundaries (Fig. 1 (a)), even considering that adjacent domains are geometrically aligned. Instead, we here consider that a pyramid is to be more suitable for the formation of single-crystal BN multilayers from well-aligned adjacent BN domains (Fig. 1 (b)). Our measurements by atomic force microscopy (AFM) show a very small bottom angle ($\beta \sim 6.73°$, Figs. 1 (c) - (e), also seen in the cross-sectional high-resolution transmission electron microscopy (HRTEM) image, Fig. S1) of the pyramid, translating into a ~ 2.8 nm/5.6 nm separation between two adjacent layers edges (Fig. 1 (c)). These two values are also added (2.8 nm + 5.6 nm, Fig. 1 (c)) when two domains meet and proceed to connect. This mechanism imposes a steric separation that effectively reduces the competitive connection of approaching BN layers, thus eliminating the possible misconnection (that by various kinds of grain boundaries, e.g., 5-7 rings [26]) among adjacent layers, or the formation of broken layer edges (see Fig. 1 (a)). The pyramid structure (Figs. 1 (f) – (j)) is thus superior in the sense that it enables a "first-meet-first-connect" mechanism and it finally leads to a continuously connecting single-crystal multilayer (CSM, Fig. 1 (k)). An AB or ABC (viz. the unidirectional stacking) -stacked BN multilayer will thus finally be obtained following this mechanism by the observation of effective layer connection at the domain boundary. This growth mode also leaves part of the pyramid, seen as the pyramid apex, as a residue non-connecting island (RNI, Fig. 1 (k)), which might also be realized as an indicator of the current mechanism. From another aspect, as the orientation of the pyramidal domains is defined by the first layer of the BN islands, to use a single-crystal metal substrate helps effectively to achieve a unidirectional orientation of the BN islands[14]. The pyramidal BN domains that grown aligned on a single-crystal metal substrate are mandatory to enable the subsequent growth of the single-crystal multilayer BN films with large area.

The above strategy is then verified successfully based on our experiment observations. It is shown that a predominant alignment of multilayer BN pyramids grown on a single-crystal Fe-Ni (111) surface is appropriate for the successful connection of the BN layers. Large-area single-crystal multilayer BN thin films with AB (Bernal) and ABC (Rhombohedral) stackings were observed. The full surface coverage as well as the single-crystal structure of the BN thin films are evidenced by various techniques.

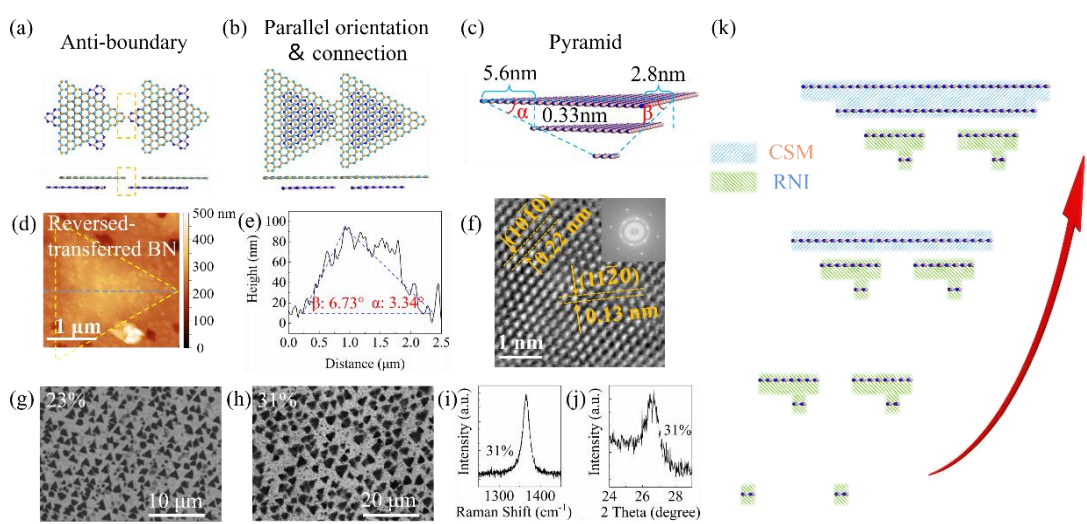

*Fig. 1 | Growth mechanism of multilayer boron nitride (BN) films on Fe-Ni (111). (a) Hexagram structure (AA' stacking) of BN domains connecting with grain boundaries. (b) Pyramidal structure (AB stacking) of BN domains connecting without grain boundary. (c) Schematic diagram of the atomic model of the pyramidal structure. (d) Atomic force microscopy of BN backside structure, and (e) height profile along the blue*

*dashed line in (d). (f) High resolution transmission electron microscope (HRTEM) image of the BN film. The inset shows the fast Fourier transform of the whole image. The d-spacings of ($10\bar{1}0$) and ($11\bar{2}0$) planes of BN are 0.22 and 0.13 nm, respectively. (g) Scanning electron microscope images of BN grown on Fe-Ni (111) with the Fe content of 23% (g), and 31% (h). (i) Raman spectrum of BN grown on Fe-Ni (111) with the Fe content of 31%. (j) X-ray diffraction pattern of BN grown on Fe-Ni (111) with the Fe content of 31%. (k) Schematic of the growth mechanism of single-crystal multilayer BN films on Fe-Ni (111) alloy film. The continuously connecting single-crystal multilayer (CSM) and residue non-connecting island (RNI) are marked with different colors.*

## Results and discussions

Single-crystal Fe-Ni alloy (111) substrates were deposited on commercial c-plane sapphire by combining magnetron sputtering and thermal evaporation (Fig. S2). Doping a Ni film with an appropriate amount of Fe improves the solubility of boron and nitrogen in the catalyst, which is beneficial for the growth of multilayer BN [27, 28]. We found a stable face-centered cubic structure of the Fe-Ni (111) thin films up to 51% Fe (Fe/(Fe+Ni) in at %) (Fig. S3) [29, 30]. According to the XRD results, the lattice constant increases slightly from 3.55 Å to 3.59 Å, which roughly matches the BN lattice. A control of the BN morphology upon increasing the Fe content in the alloy was observed (Figs. S4, S5). Basically, increasing the Fe content enhances the BN growth by increasing the density of nucleation sites, the BN domain size, and the BN coverage and/or thickness, as evidenced by scanning electron microscopy (SEM), and also by X-ray diffraction (XRD) and Raman spectroscopy (Figs. S4 (a) - (f), (h), (i), and Fig. S6). Transmission electron microscopy (TEM) further confirmed the formation of a well-crystallized BN lattice (Fig. 1 (f)). At medium Fe content (23 % typically, Fig. 1 (g)), BN pyramidal domains with well-shaped triangular basis were seen, with domain size of ~ 1.34 μm and an areal density of 0.4 $\mu m^{-2}$ (Figs. S4 (g), (j)). A statistical analysis shows that 83.5% of the BN domains are unidirectionally aligned (for 31% Fe-Ni sample, some local statistical results are shown in Fig. S7, see also Fig. S4 (j) for statistics on other samples). Comparing with conventional ~ 1:1 ratio (50% unidirectional alignment) of oppositely aligned BN domains grown by the CVD method observed on single-crystal substrates [31 - 33], we conclude on the pronounced unidirectional alignment in our samples, which is highly beneficial for the effective connection of BN layers from adjacent BN domains (as displayed in Figs. 1 (b), (k)).

The interconnection of the BN multilayers among adjacent BN pyramids and the formation of a continuous BN layer (the CSM region) are first evidenced by an enhanced resistance to oxidation of the Fe-Ni

(111) surface. We compared the surface morphology of as-deposited samples and samples stored in ambient air for a long period (~ 1 year) by AFM. For the as-deposited samples on 23% Fe-Ni, clean surfaces with clear BN triangles were seen on the metal alloy surface (Fig. 2 (a)). After ~ 1-year storage, it was profoundly oxidized at regions in-between BN triangles by showing many abrupt particles with a size of several tens to hundreds of nanometers (Fig. 2 (b)). We note also that the regions covered by BN triangular domains were still well-protected without any oxidation. This reveals that the 23 % Fe-Ni (111) substrate is not fully covered by multilayer BN. In contrast, the samples with higher Fe content, e.g., 31% Fe, were not oxidized as seen from the smooth surface structure seen in both as-deposited and long-time stored (~ 1 year) samples (Figs. 2 (c), (d)), telling that the Fe-Ni (111) surface is completely protected by a full BN coverage. The same situation was also more evidently observed on the surface of samples with still higher (41% and 51%) Fe content (Figs. S8 (a) - (d)). However, it is worth mentioning that the BN domains of the inverted pyramid structure grow by sinking into the metal substrate, so the AFM can only obtain part of the surface information. The inverted pyramid structure or the layer connection between adjacent domains might not be resolved by AFM. To further showcase the large-area coverage and uniformity of the grown multilayer BN film, wafer-scale BN (~ 2 inches) on the 51% Fe sample was

12 / 62

transferred onto $SiO_2$/Si (300 nm $SiO_2$) (Fig. S9). Intact and homogeneous BN multilayers can be visually observed, either from optical photograph (Fig. S9 (d)) or from low-magnification microscope (Fig. S9 (e)).

X-ray photoelectron spectroscopy (XPS) was also used to study the BN thin films upon increasing Fe content (Fig. S10). Two samples were measured for the moment, in which the 31% Fe sample is estimated to have lower surface coverage comparing with the 51% Fe one. Considering the analyze depth of the XPS technique and long storage time, we anticipate Ni, Fe, B, N, O, C elements to be observed on both samples, while the 51% Fe samples should show evidently lower O/N ratio (or O/Ni ratio). The XPS results fit well to this anticipation (as seen in Fig. S10 (b)), indicating the increased coverage in high Fe content samples. The observation of metal oxide related XPS peak in the 31% Fe sample (while not observed in the 51% Fe sample) also verifies the change of BN coverage. This observation is also in well accordance with the AFM results.

In addition to the full surface protection by BN layers at high Fe content samples, we clearly observed CSM and RNI regions, which are the two types of BN deposits involved in the growth mechanism shown in Fig. 1 (k). These regions are easily seen in Figs. 1 (g), (h) and Figs. S4 (e), (f), and are also observable with an optical microscope (Figs. 2 (e) - (h)) in the sample that was transferred onto $SiO_2$(285 nm)/Si substrate. Such structure was also verified by AFM (Figs. 2 (a) - (d) and Fig. S8). The size of the RNI regions varies upon increasing Fe content: while the 41% Fe-sample shows an RNI size reduction compared to 31% Fe (Figs. S4 (e), (j)), probably due to an enhanced layer connection that consumes more layers from the pyramid bottom side, the increased RNI size in the 51% Fe-sample (Figs. S4 (f), (j)) indicates an enhanced growth of the pyramids under this high Fe content. A selected area electron diffraction (SAED) analysis performed around and inside the RNI region (Fig. 2 (i)) indicates that the CSM and RNI regions are well aligned without any misorientation, again supporting the mechanism showing in Fig. 1 (k), and further discussed below.

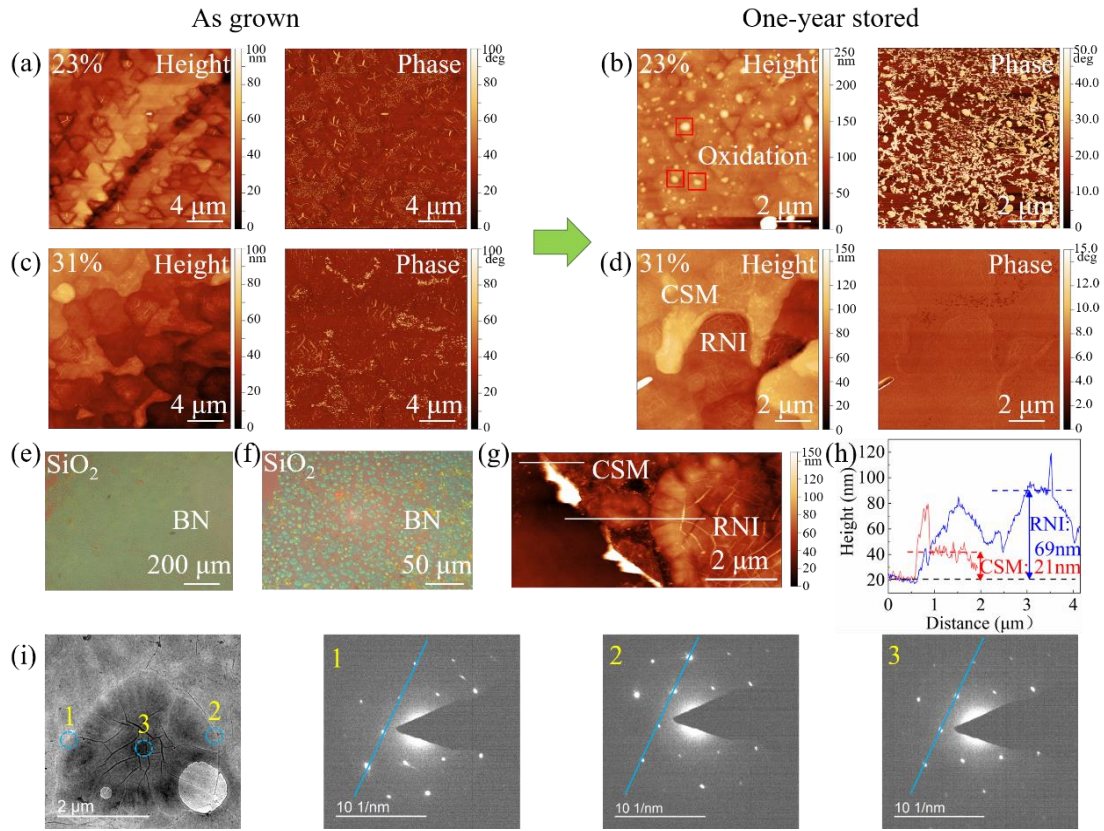

*Fig. 2 | Atomic force microscopy images of transferred multilayer BN films grown on alloy film: as-deposited samples (a), samples stored for 1 year (b), for a Fe content of 23% in the Fe-Ni substrate; as-deposited samples (c), and samples stored for 1 year (d), for a Fe content of 31%. (e), (f) Optical microscope images of the transferred BN films on SiO$_2$/Si substrates. AFM image of the transferred BN film (g), and height profiles (h) along the lines shown in (g). The thickness of connecting single-crystal multilayer (CSM) and residue non-connecting island (RNI) are 21 nm and 69 nm, respectively. (i) TEM morphology images and selected area electron diffraction patterns (1~3) at different positions indicated in the triangular BN domain.*

We performed detailed cross-sectional HRTEM observations in order to confirm the effective edge connection as predicted in Fig. 1(k). A location of the connection site (selected randomly under SEM observation) was accurately confirmed via the focused ion beam (FIB) process (see Figs. 3 (a) - (e)) and the layer connection is confirmed directly by analyzing the edge-type dislocation distribution with Burgers vector $b$= [*0001*]. The projected areal density (count to be edge-type dislocation. It is worth to note that one screw-type might count to be more than one edge-type under such geometry [34], however it does not bias the comparison) is found to be 0.047/nm$^2$ at the site (Fig. 3 (e)) where the two BN domains connect, comparing with the density of 0.036 ~ 0.071/nm$^2$ (averaged at 0.057/nm$^2$, see Figs. 3 (f) - (h) and Fig. S11) that is observed in other randomly chosen regions inside individual BN domains, as shown in Fig. S11. It is thus concluded that the dislocation density at the site to connect is comparable or even lower to that in individual domain regions, which unambiguously confirms the effective layer connection between two adjacent multilayer BN domains.

We also made another FIB sample in another randomly chosen site (results shown in Figs. S12 and S13). The dislocation density at the site to connect (0.048/nm$^2$) is also found identical to the dislocation density (0.050/nm$^2$) in individual domain regions. To this end, it is thus quite convincing to conclude on the effective edge connection in the multilayer

BN domains.

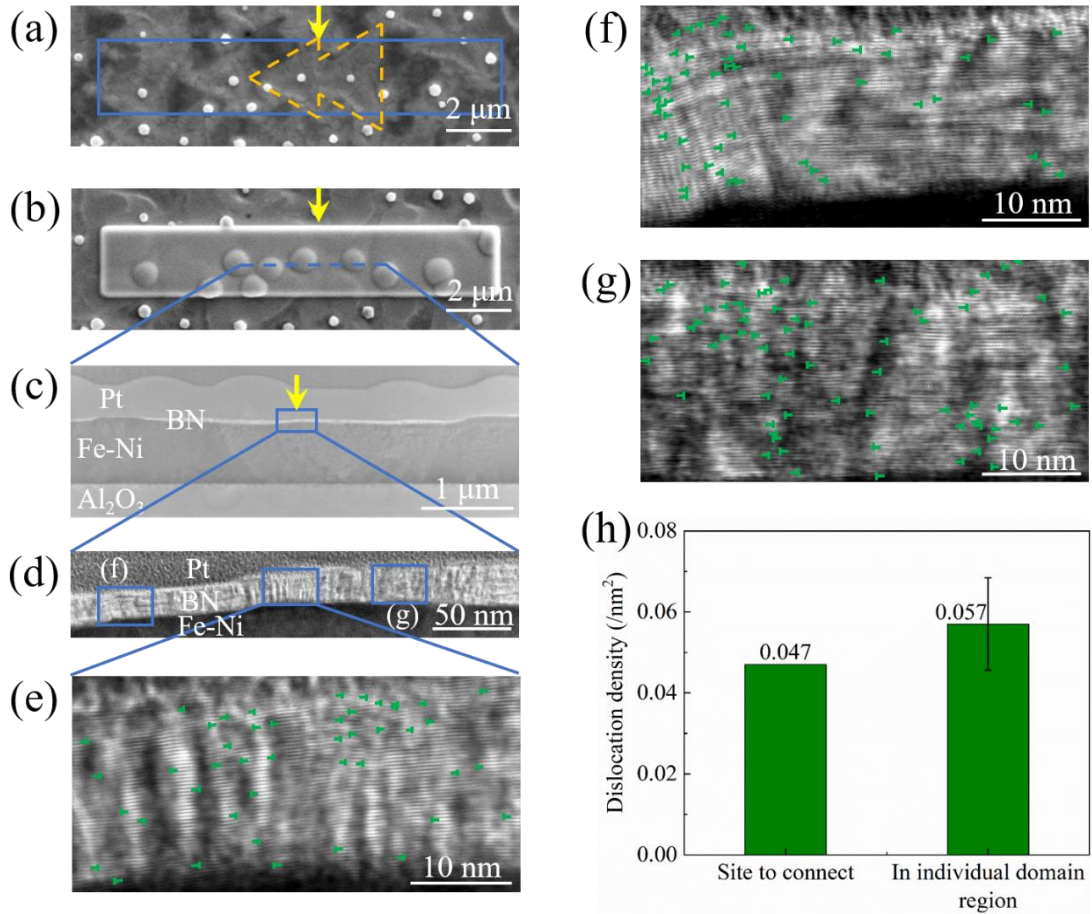

*Fig. 3 | Focused ion beam (FIB) preparation process and statistical analysis of dislocation density of boron nitride (BN) multilayers. (a) Scanning electron microscope (SEM) image of the area chosen for the cross-sectional transmission electron microscopy (TEM) sample, in which the bule line frames the area for Pt deposition, the orange line frames the two triangular BN domains for the cross-sectional observation, and the yellow arrow mark the site to connect of two triangular domains. (b) The SEM image after Pt deposition. (c-e) Cross-sectional TEM image of the site to connect at increasing magnifications. The dislocations (b= [0001])*

*are marked with the green symbols, as also indicated in (f) and (g). (f), (g) The cross-sectional high-resolution transmission electron microscopy (HRTEM) images in individual domain region, and the relative positions are shown in (e). (h) Statistical histogram of dislocation density (**b**= [0001]) of the site to connect and inside individual BN domain region. The error bar is the standard deviation.*

We further investigated the alignment and stacking of the synthesized BN multilayers with detailed TEM and SAED measurements. An area of ~ 2 mm × 2.2 mm was used for SAED survey (Fig. 4 (a), also see all of the SAED results in Fig. S14), in which 67 points are selected randomly for a detailed inspection. The parallel blue lines in the different SAED patterns prove that the BN orientation is the same over a large area (note that oppositely-aligned pyramids are not distinguishable using SAED). SAED surveys performed at the CSM regions give also the same orientation (Fig. S15 (a)).

The existence of anti-domains naturally creates grain boundaries that then constitute weak points and induce film fractures. We thus anticipate the presence of such anti-domain boundaries in our sample, considering the non-perfect ~ 80% unidirectional orientation of the RNI. However, our detailed TEM observations did not resolve any evident crack lines in the 2 mm × 2.2 mm region. It was found that the whole film is crack-free with just some regions displaying some contrast variation (Fig. S16). We could not resolve if these structures are anti-domain boundaries. It is suggested that the anti-domains might be effectively merged via grain boundary structures [33, 35, 36].

The large-area orientation and the coverage of the multilayer BN films was also verified by large scale low energy electron diffractions (LEED) survey over a region of ~3 mm × 2 mm. We observed sharp diffraction spots on the 51% Fe sample (Fig. S17) in total of 116 spots. All these diffraction patterns are completely consistent, confirming that the crystalline lattice of the BN multilayers is well aligned along the same direction. This observation intensified the observation using SAED (Fig. 4 (a) and Fig. S14). An intensity variation of the LEED spots from site to site could be possibly raised by the thickness variation across the BN domains, e.g., the CSM and RNI regions (Fig. 1 (k)). As a comparison, however, we observed vague diffraction on the 31% Fe sample (no meaningful diffraction was recorded), which could be due to the lower surface BN coverage and also the rough sample surface (see the AFM image in Figs. 2 (c) (d)).

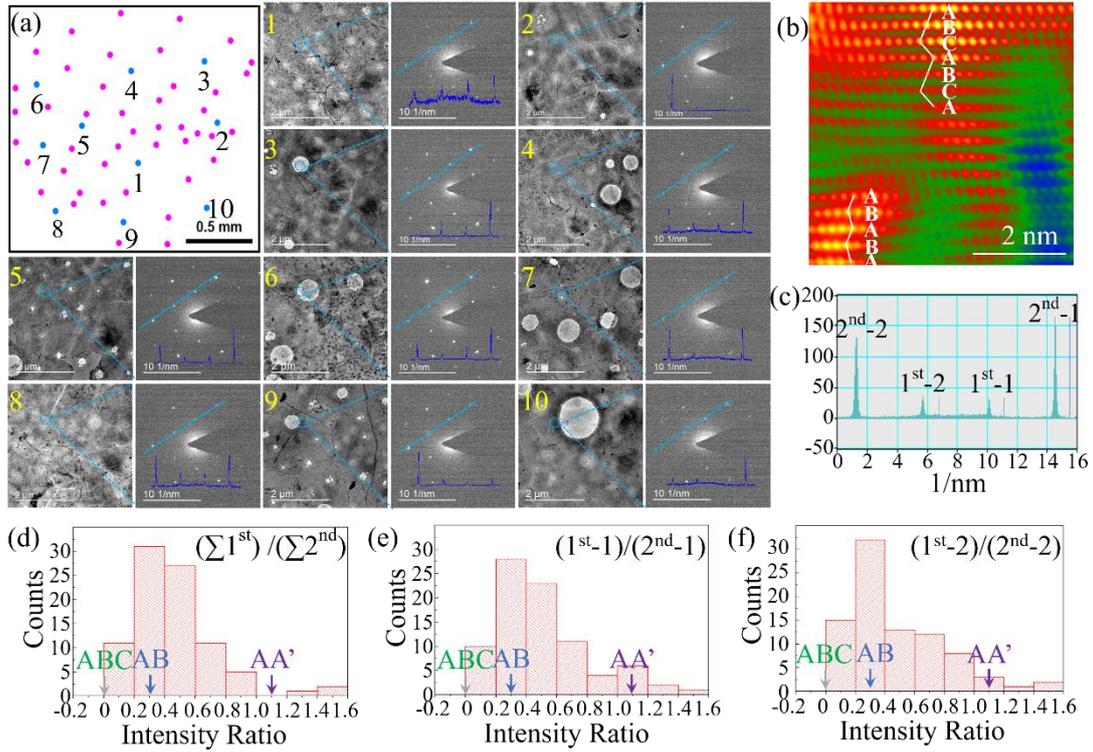

*Fig. 4 | TEM and selected area electron diffraction (SAED) of the synthesized BN films. (a) Global map of the randomly selected points in a large area, 2 mm×2.2 mm. TEM images and SAED patterns for ten points (labelled from 1 to 10). (b) The cross-sectional high resolution transmission electron microscope image of BN films. There are both AB and ABC stacking, but no AA'. (c) Typical electron diffraction intensity distribution. (d)-(f) Statistical analysis of the stacking orders of BN with three different criteria.*

The stacking of the BN films was examined by comparing the intensity of first-order and second-order diffraction peaks in SAED (Fig. 4 (a)) and HRTEM (Fig. 4 (b)), in order to verify the AB or ABC parallel-type stackings, as given by the mechanism shown in Fig. 1 (k).

Theoretically, the ratio of first- and second-order diffraction peaks is calculated to be 0.3 for AB stacking, 1.1 for AA', and 0 for ABC [23]. Our typical electron diffraction intensity distribution is shown in Fig. 4 (c). Due to the intensity variations in our SAED measurements (possibly due to the film deformation that was caused by the non-ideal wet-transfer process and rough TEM grid surface), we calculated in three different ways: (i) $(\sum 1^{st})/(\sum 2^{nd})$, (ii) $(1^{st}-1)/(2^{nd}-1)$, and (iii) $(1^{st}-2)/(2^{nd}-2)$ (Figs. 4 (d) - (f)). We see a constant trend that the diffraction intensity of first order (inner peaks) is always lower than that of second order (outer peaks) (Fig. 4 (c)) regardless of the calculation methods. Thus, the AA' stacking is largely excluded. Our detailed statistical analysis (data from 54 random positions) indicates that the BN multilayer is dominated by the AB stacking, with a part of ABC stacking, while the AA' phase is unlikely (Fig. 4). The cross-sectional HRTEM results also confirm the conclusion of our statistical analysis (Fig. 4 (b)). According to our detailed HRTEM observations (see Fig. 4 (b)), both AB and ABC stackings are observed, while AA' stacking is not. Such an observation meets the estimation following the proposed growth mechanism and observed pyramidal structures (Figs. 1 (b), (c), (k)).

The optical properties of multilayer BN (on Fe-Ni single crystal substrate with 31% Fe) were further studied by spatially-resolved photoluminescence (PL) microscopy in the deep ultraviolet, performed at cryogenic temperatures (T ~ 6 K) [37]. Figs. 5 (a), (b), (c) display the PL intensity maps recorded on the exact same region of the BN multilayer, but integrated in three different spectral domains represented in red on Figs. 5 (d), (e), (f). The three maps are quite similar, and they display three hot spots that we attribute to BN pyramids. Because the thickness of the deposited BN is the largest at the RNI pyramidal islands (Fig. 1 (k)), a brighter PL signal is expected at these positions. The number and the few µm-extension of the PL hot spots in Figs .4 (a), (b), (c) are fully consistent with the density and size of the RNI islands characterized by AFM and SEM experiments in the sample (Fig. 2 and Fig. S4). Second harmonic generation measurements (Fig. S18) further confirm the optical identification of the RNI pyramidal domains using our confocal scanning microscope operating in the deep ultraviolet.

The PL spectrum of the brightest spot, averaged over the red square in Figs. 5 (a), (b), (c), is displayed in red line in Figs. 5 (d), (e), (f), for the three corresponding spectral ranges under investigation. For the sake of the comparison with the optical response of well-identified BN polytypes [38 - 43], we also plot the typical PL spectra of hBN in the AA' stacking (black line in Figs. 5 (d), (e), (f)) and Bernal BN (bBN) in the AB stacking (blue line in Figs. 5 (d), (e), (f)) [40]. In contrast to the AA' and AB stackings, there is so far no characterization by PL spectroscopy of BN monocrystals in the ABC stacking. Even though the PL signal of ABC-stacked BN was identified as lying in the 5.3-5.5 eV range [39], it spectrally overlaps with the emission of excitons localized at stacking faults, thus precluding here any investigation of BN in the ABC stacking by PL spectroscopy. We thus focus in the following on the comparison of the PL signal of our CVD-grown BN with the optical signatures of BN in the AA' and AB stackings.

Figs. 5 (a), (d) focus on the low energy range, around 4.1 eV (wavelength~300 nm), corresponding to the spectral domain of deep level emission in BN. The PL spectrum shows an asymmetric line centered at 4.14 eV, which is attributed to the well-known so-called "4 eV" defect in BN. Recent measurements have shown that the PL spectrum of this defect depends on the stacking order of the BN host matrix [41, 42]. While the zero-phonon line (ZPL) of this defect is typically found at 4.09 eV in pure AA' stacking (hBN), it is blue-shifted with two components at energies 4.14 eV and 4.16 eV for the AB stacking (bBN) (Fig. 5 (d)) [42]. The lack of any line at 4.09 eV indicates the absence of the AA' stacking in our CVD-grown multilayer BN. Conversely, the PL signal centered at 4.14 eV matches with the blue-shifted ZPLs in the Bernal phase of BN in the AB stacking [42]. The broadening of the PL line centered at 4.14 eV (red line, Fig. 5 (d)) does not allow to investigate the presence of different ZPL energies in our CVD-grown sample. Nonetheless, the 4.14 eV-energy of the broad ZPL is a first indication in PL spectroscopy of an AB stacking of our multilayer BN.

The spectral range around 5.5 eV (wavelength ~ 225 nm) corresponds to the shallow defect emission band (Figs. 5 (b), (e)), which originates from extended defects and stacking faults. The PL spectrum of our CVD-grown BN shows two broad bands centered at 5.37 eV and 5.53 eV (red line, Fig. 5 (e)), that are again spectrally aligned with the ones recorded in bBN (blue line, Fig. 5 (e)) [40], but with a different intensity ratio. Noticeably, there is a local minimum of the PL spectrum in our CVD-grown BN at the energy of ~ 5.47 eV, for which the PL signal displays its absolute maximum in hBN, thus confirming the absence of the AA' order and the existence of a predominant AB stacking.

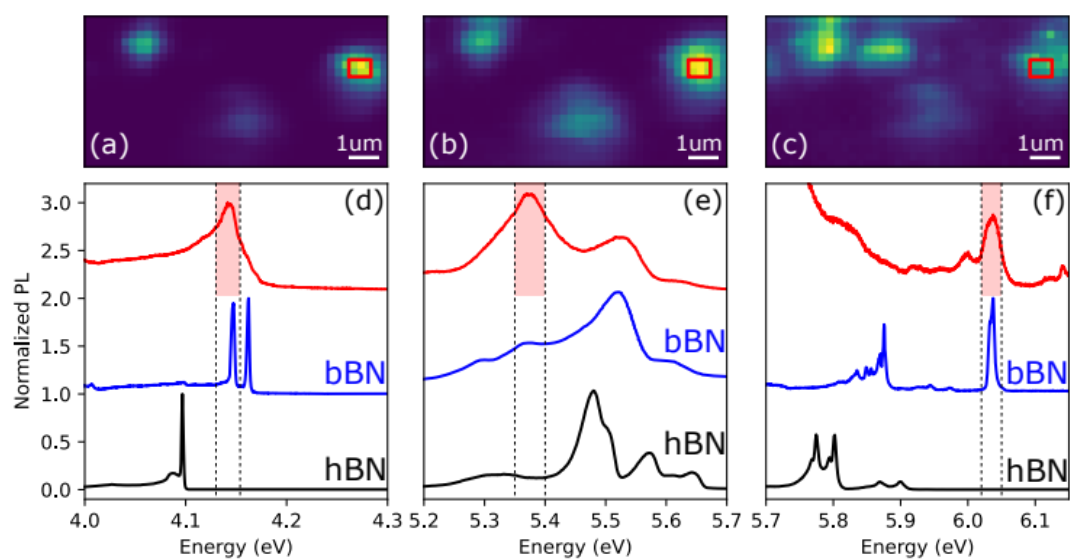

*Fig. 5 | Spatially-resolved photoluminescence (PL) spectroscopy in the deep ultraviolet for multilayer BN grown on the 31% Fe-substrate. (a), (b), (c) PL intensity maps recorded on the exact same region of the BN*

*multilayer, but integrated in three different spectral domains represented in red on panels (d), (e), (f). (d), (e), (f) PL spectra spatially integrated in the region indicated by the red square in panels (a), (b), and (c) (red line), Bernal BN (bBN) in the AB stacking (blue line), hexagonal boron nitride (hBN) in the AA' stacking (black line).*

Finally, Figs. 5 (c), (f) address the intrinsic band edge emission in the 6 eV-range (wavelength ~ 200 nm). A unique signature of the Bernal phase of BN is the existence of an emission line around ~ 6.035 eV (Fig. 5 (f)), in a spectral domain where strictly no luminescence is detected in the case of hBN [40]. The detection of an emission line around 6.035 eV in the present multilayer BN (red line, Fig. 5 (f)) is an additional signature for the preferential AB stacking order in our CVD-grown BN. Interestingly, the PL intensity map around 6 eV (Fig. 5 (c)) is no longer dominated by only three hot spots as for the deep (Fig. 5 (a)) and extended (Fig. 5 (b)) defects. This observation brings two important information. First, most of the defects are located in the RNI domains of multilayer BN (Fig. 1 (k)). Second, there is intrinsic band edge emission in between the RNI islands, thus demonstrating the quality of the CSM layers in the Bernal phase of BN.

We thus conclude that, in the three investigated spectral domains of emission, our CVD-grown multilayer BN features the specific optical response of the Bernal phase of BN in the AB stacking, in agreement with the structural characterizations concluding to the presence of both AB and ABC stackings (Fig. 4), the investigation of the ABC phase being beyond reach here because of the spectral overlap of its PL signal with the stacking faults emission.

Looking more deeply into our BN multilayers by complementary tools, conductive atomic force microscopy (cAFM) was used to map the leakage current distribution of the multilayer BN (Figs. 6 (a) – (c), also Fig. S19 and Fig. S20). The areas with the largest current levels (Figs. 6 (b), (c)) appear to be the thinner CSM regions, while the regions having lower current level are realized to be thicker RNI regions (to be interpreted by combining the morphologies in Figs. 6 (a), (b)). One also resolves triangular features with CSM and RNI regions as indicated both in the morphology and current maps (Fig. S19), in roughly agreement with the growth mechanism discussed above (Fig. 1 (k)).

No evident weak leakage lines were detected among the CSM regions, consistently with an effective connection of the BN layers. This property ensures excellent screening of current leakage via defects in BN layers. Considering a typical leakage current on the order of ~ 10 nA under a bias of 6 V (electrical field ~ $10^8$ V/cm, assuming a ~ 60 nm BN thickness), the BN multilayer appears to be of considerably high quality. More current mapping performed in randomly chosen areas over the sample surface (Fig. S19) shows similar features. No evident weak leakage lines among CSM regions were observed. Low current level is observed even in regions that are tentatively attributed to anti-domains (Fig. S20). This might benefit from the formation of non-coincident anti-domain boundary that also works for effective screening of possible leakage paths (Fig. S20 (c)).

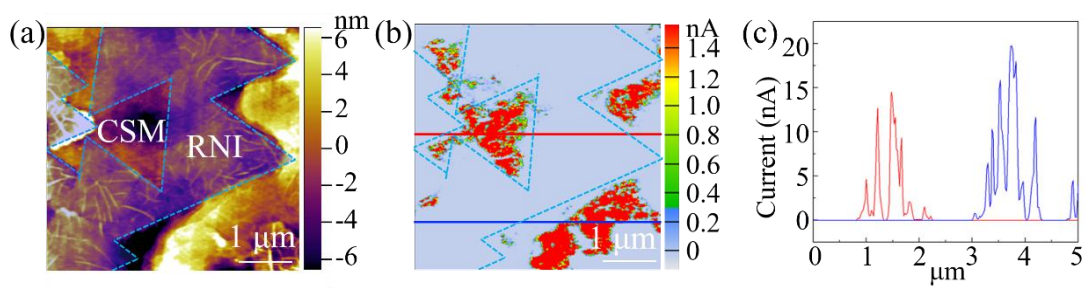

*Fig. 6 | Conductive atomic force microscopy (cAFM) measurements of multilayer BN, under the bias voltage of 6 V. (a) cAFM height image. (b) cAFM current mapping. The residue non-connecting islands (RNI) are marked with the false-color (blue), with small leakage current (pA levels), and the other regions correspond to the continuously connecting single-crystal multilayer (CSM). (c) Current profiles along the two lines indicated in panel (b).*

## Conclusion

In summary, multilayer BN films were successfully grown on single-crystal Fe-Ni alloy by CVD, with 83% alignment of the pyramidal BN

nucleation islands and a subsequent successful layer-by-layer connection leading to a continuous multilayer film. It is realized that the single crystal metal helps on achieving the unidirectional BN domain orientation; Fe is introduced in order to promote the multilayer growth; and the Fe content is to be designed and controlled carefully in order to maintain the phase stability of the metal single crystal. The Fe-Ni (111) single-crystal film, which has a stable face-centered cubic structure up to 51% Fe, and has stronger catalytic ability for the growth of BN multilayers, favors the alignment of BN domains in this novel growth mechanism following a "first meet-first connect" process of the stacked parallel layers. Our detailed TEM observation verified this mechanism by showing effective connection between BN domains. A statistical analysis of selective area electron diffraction and the cross-sectional HRTEM diagrams reveal the AB and ABC stacking of our multilayer BN films obtained by CVD, in agreement with the expected natural exclusion of the anti-parallel AA' stacking in the present growth mechanism. AB and ABC stackings mean the fabrication of BN multilayers in the Bernal and Rhombohedral phase, in contrast to the most thermodynamically stable AA' stacking routinely obtained in bulk hBN crystals. Spatially-resolved photoluminescence spectroscopy in the deep ultraviolet reveals that the CVD-grown multilayer BN features the specific optical response of the Bernal phase of BN in the AB stacking, fully confirming the structural

28 / 62

characterizations by electron diffraction and cross-sectional HRTEM.

We believe that higher unidirectional alignment of domains on single-crystal Fe-Ni (111) could be achieved in the future with this CVD method, for a boundary-free single-crystal of multilayer BN film of large thickness, by careful control of the single-crystal metal surface structures and growth parameters.

**Acknowledgements**

This work is supported by the National Natural Science Foundation of China (Grant Nos. 61504030, 61370042, and 61006011). This work was financially supported by the BONASPES project (No. ANR-19-CE30-0007), the ZEOLIGHT project (No. ANR-19-CE08-0016) and Université de Montpellier.

# References

[1] Tran, T. T.; Bray, K.; Ford, M. J.; Toth, M.; Aharonovich, I. Quantum emission from hexagonal boron nitride monolayers. Nature Nanotech 2016, 11, 37–41.

[2] Bourrellier, R.; Meuret, S.; Tararan, A.; Stephan, O.; Kociak, M.; Tizei, L. H. G.; Zobelli, A. Bright UV Single Photon Emission at Point Defects in h-BN. Nano Lett. 2016, 16, 7, 4317–4321.

[3] Dean, C. R.; Young, A. F.; Meric, I.; Lee, C.; Wang, L.; Sorgenfrei, S.; Watanabe, K.; Taniguchi, T.; Kim, P.; Shepard, K. L.; Hone, J. Boron nitride substrates for high-quality graphene electronics. Nature Nanotech 2010, 5, 722–726.

[4] Fukamachi, S.; Solis-Fernandez, P.; Kawahara, K.; Tanaka, D.; Otake, T.; Lin, Y. C.; Suenaga, K.; Ago, H. Large-area synthesis and transfer of multilayer hexagonal boron nitride for enhanced graphene device arrays. Nat Electron 2023, 6, 126–136.

[5] Lu, Z. J.; Zhu, M. J.; Liu, Y. F.; Zhang, G. H.; Tan, Z. Q.; Li, X. T.; Xu, S. S.; Wang, L.; Dou, R. F.; Wang, B.; Yao, Y.; Zhang, Z. Y.; Dong, J. C.; Cheng, Z. H.; Chen, S. S. Low-Temperature Synthesis of Boron Nitride as a Large-Scale Passivation and Protection Layer for Two-Dimensional Materials and High-Performance Devices. ACS Appl. Mater. Interfaces 2022, 14, 22, 25984–25992.

[6] Liu, Z.; Gong, Y. J.; Zhou, W.; Ma, L. L.; Yu, J. J.; Idrobo, J. C.; Jung, J.; MacDonald, A. H.; Vajtai, R.; Lou, J.; Ajayan, P. M. Ultrathin high-temperature oxidation-resistant coatings of hexagonal boron nitride. Nat Commun 2013, 4, 2541.

[7] Chilkoor, G.; Jawaharraj, K.; Vemuri, B.; Kutana, A.; Tripathi, M.; Kota, D.; Arif, T.; Filleter, T.; Dalton, A. B.; Yakobson, B. I.; Meyyappan, M.; Rahman, M. M.; Ajayan, P. M.; Gadhamshetty, V. Hexagonal Boron Nitride for Sulfur Corrosion Inhibition. ACS Nano 2020, 14, 11, 14809–14819.

[8] Kubota, Y.; Watanabe, K.; Tsuda, O.; Taniguchi, T. Deep ultraviolet light-emitting hexagonal boron nitride synthesized at atmospheric pressure. Science 2007, 317, 932-934.

[9] Maity, A.; Grenadier, S. J.; Li, J.; Lin, J. Y.; Jiang, H. X. High efficiency hexagonal boron nitride neutron detectors with 1 cm2 detection areas. Appl. Phys. Lett. 2020, 116, 142102.

[10] Kobayashi, Y.; Kumakura, K.; Akasaka, T.; Yamamoto, H.; Makimoto, T. Layered boron nitride as a release layer for mechanical transfer of GaN-based devices. Nature 2012, 484, 223-227.

[11] Park, J. H.; Yang, X.; Lee, J. Y.; Park, M. D.; Bae, S. Y.; Pristovsek, M.; Amano, H.; Lee, D. S. The stability of graphene and boron nitride for III-nitride epitaxy and post-growth exfoliation. Chem. Sci. 2021, 12, 7713-7719.

[12] Vuong, P.; Moudakir, T. Gujrati, R.; Srivastava, A.; Ottapilakkal, V.; Gautier, S.; Voss, P. L.; Sundaram, S.; Salvestrini, J. P.; Ougazzaden, A. Scaling up of Growth, Fabrication, and Device Transfer Process for GaN-based LEDs on H-BN Templates to 6-inch Sapphire Substrates. Adv. Mater. Technol. 2023, 2300600.

[13] Wang, L.; Xu, X. Z.; Zhang, L. N.; Qiao, R. X.; Wu, M. H.; Wang, Z. C.; Zhang, S.; Liang, J.; Zhang, Z. H.; Zhang, Z. B.; Chen, W.; Xie, X. D.; Zong, J. Y.; Shan, Y. W.; Guo, Y.; Willinger, M.; Wu, H.; Li, Q. Y.; Wang, W. L.; Gao, P.; Wu, S. W.; Zhang, Y.; Jiang, Y.; Yu, D. P.; Wang, E. G.; Bai, X. D.; Wang, Z. J.; Ding, F.; Liu, K. H. Epitaxial growth of a 100-square-centimetre single-crystal hexagonal boron nitride monolayer on copper. Nature 2019, 570, 91–95.

[14] Chen, T. A.; Chuu, C. P.; Tseng, C. C.; Wen, C. K.; Wong, H. S. P.; Pan, S. Y.; Li, R. T.; Chao, T. A.; Chueh, W. C.; Zhang, Y. F.; Fu, Q.; Yakobson, B. I.; Chang, W. H.; Li, L. J. Wafer-scale single-crystal hexagonal boron nitride monolayers on Cu (111). Nature 2020, 579, 219–223.

[15] Knobloch, T.; Illarionov, Y. Y.; Ducry, F.; Schleich, C.; Wachter, S.; Watanabe, K.; Taniguchi, T.; Mueller, T.; Waltl, M.; Lanza, M. R.; Vexler, M. I.; Luisier, M.; Grasser, T. The performance limits of hexagonal boron nitride as an insulator for scaled CMOS devices based on two-dimensional materials. Nat Electron 2021, 4, 98–108.

[16] Kim, S. M.; Hsu, A.; Park, M. H.; Chae, S. H.; Yun, S. J.; Lee, J. S.; Cho, D. H.; Fang, W. J.; Lee, C.; Palacios, T.; Dresselhaus, M.; Kim, K. K.; Lee, Y. H.; Kong, J. Synthesis of large-area multilayer hexagonal boron nitride for high material performance. Nat Commun 2015, 6, 8662.

[17] Park, J. H.; Choi, S. H.; Zhao, J.; Song, S.; Yang, W.; Kim, S. M.; Kim, K. K.; Lee, Y. H. Thickness-controlled multilayer hexagonal boron nitride film prepared by plasma-enhanced chemical vapor deposition. Current Applied Physics 2016, 16, 1229-1235.

[18] Shi, Z. Y.; Wang, X. J.; Li, Q. T.; Yang, P.; Lu, G. Y.; Jiang, R.; Wang, H. S.; Zhang, C.; Cong, C. X.; Liu, Z.; Wu, T. R.; Wang, H. M.; Yu, Q. K.; Xie, X. M. Vapor–liquid–solid growth of large-area multilayer hexagonal boron nitride on dielectric substrates. Nat Commun 2020, 11, 849.

[19] Oshima, C.; Nagashima, A. Ultra-thin epitaxial films of graphite and hexagonal boron nitride on solid surfaces. J. Phys.: Condens. Matter 1997, 9, 1.

[20] Ma, K. Y.; Zhang, L.; Jin, S.; Wang, Y.; Yoon, S. I.; Hwang, H.; Oh, J.; Jeong, D.; Wang, M.; Chatterjee, S.; Kim, G.; Jang, A. R.; Yang, J.; Ryu, S.; Jeong, H. Y.; Ruoff, R. S.; Chhowalla, M.; Ding, F.; Shin, H. S. Epitaxial single-crystal hexagonal boron nitride multilayers on Ni (111). Nature 2022, 606, 88–93.

[21] Uchida, Y.; Kawahara, K.; Fukamachi, S.; Ago, H. Chemical Vapor Deposition Growth of Uniform Multilayer Hexagonal Boron Nitride Driven by Structural Transformation of a Metal Thin Film. ACS Appl. Electron. Mater. 2020, 2, 10, 3270–3278.

[22] Uchida, Y.; Nakandakari, S.; Kawahara, K.; Yamasaki, S.; Mitsuhara, M.; Ago, H. Controlled growth of large-area uniform multilayer hexagonal boron nitride as an effective 2D substrate. ACS Nano 2018, 12, 6236–6244.

[23] Gilbert, S. M.; Pham, T.; Dogan, M.; Oh, S.; Shevitski, B.; Schumm, G.; Liu, S.; Ercius, P.; Aloni, S.; Cohen, M. L.; Zettl, A. Alternative stacking sequences in hexagonal boron nitride. 2D Mater. 2019, 6, 021006.

[24] Babenko, V.; Fan, Y.; Veigang-Radulescu, V. P.; Brennan, B.; Pollard, A. J.; Burton, O.; Alexander-Webber, J. A.; Weatherup, R. S.; Canto, B.; Otto, M.; Neumaier, D.; Hofmann, S. Oxidising and carburising catalyst conditioning for the controlled growth and transfer of large crystal monolayer hexagonal boron nitride. 2D Mater. 2020, 7, 024005.

[25] Chen, T. A.; Chuu, C. P.; Tseng, C. C.; Wen, C. K.; Wong, H. S. P.; Pan, S. Y.; Li, R. T.; Chao, T. A.; Chueh, W. C.; Zhang, Y. F.; Fu, Q.; Yakobson, B. I.; Chang, W. H.; Li, L. J. Chemical vapor deposition

growth of large single-crystal mono-, bi-, tri-layer hexagonal boron nitride and their interlayer stacking. ACS Nano 2017, 11, 12057–12066.

[26] Zhang, J. J.; Sun, R.; Ruan, D. L.; Zhang, M.; Li, Y. X.; Zhang, K.; Cheng, F. L.; Wang, Z. C.; Wang, Z. M. Point defects in two-dimensional hexagonal boron nitride: A perspective. Journal of Applied Physics 2020, 128, 100902.

[27] Liu, S.; He, R.; Ye, Z. P.; Du, X. Z.; Lin, J. Y.; Jiang, H. X.; Liu, B.; Edgar, J. H. Large-Scale Growth of High-Quality Hexagonal Boron Nitride Crystals at Atmospheric Pressure from an Fe–Cr Flux. Cryst. Growth Des. 2017, 17, 9, 4932–4935.

[28] Qi, J. J.; Ma, C. J.; Guo, Q. L.; Ma, C. J.; Zhang, Z. B.; Liu, F.; Shi, X. P.; Wang, L.; Xue, M. S.; Wu, M. H.; Gao, P.; Hong, H.; Wang, X. Q.; Wang, E. E.; Liu, C.; Liu, K. H. Stacking-Controlled Growth of rBN Crystalline Films with High Nonlinear Optical Conversion Efficiency up to 1%. Adv. Mater. 2023, 2303122.

[29] Vernyhora, I. V.; Tatarenko, V. A.; Bokoch, S. M. Thermodynamics of f.c.c.-Ni–Fe Alloys in a Static Applied Magnetic Field. International Scholarly Research Notices 2012, 2012, 917836, 11.

[30] Chang, W. S.; Wei, Y.; Guo, J. M.; He, F. J. Thermal Stability of Ni-Fe Alloy Foils Continuously Electrodeposited in a Fluorborate Bath. Open Journal of Metal, 2012, 2, 18-23.

[31] Uchida, Y.; Iwaizako, T.; Mizuno, S.; Tsuji, M.; Ago, H. Epitaxial chemical vapor deposition growth of monolayer hexagonal boron nitride on a Cu(111)/sapphire substrate. Phys. Chem. Chem. Phys. 2017, 19, 8230—8235.

[32] Taslim, A. B.; Nakajima, H.; Lin, Y. C.; Uchida, Y.; Kawahara, K.; Okazaki, T.; Suenaga, K.; Hibino, H. Ago, H. Synthesis of sub-millimeter single-crystal grains of aligned hexagonal boron nitride on an epitaxial Ni film. Nanoscale, 2019, 11, 14668–14675.

[33] Yin, J.; Liu, X. F.; Lu, W.; Li, J. D.; Cao, Y. Z.; Li, Y.; Xu, Y.; Li, X. M.; Zhou, J.; Jin, C. H.; Guo, W. L. Aligned growth of hexagonal boron nitride monolayer on germanium. Small 2015,11, 5375–5380.

[34] Yang, H.; Lozano, J. G.; Pennycook, T. J.; Jones, L.; Hirsch, P. B.; Nellist, P. D. Imaging screw dislocations at atomic resolution by aberration-corrected electron optical sectioning. Nat Commun 2015, 6, 7266.

[35] Ren, X. B.; Dong, J. C.; Yang, P.; Li, J. D.; Lu, G. Y.; Wu, T. R.; Wang, H. M.; Guo, W. L.; Zhang, Z.; Ding, F.; Jin, C. H. Grain boundaries in chemical-vapor-deposited atomically thin hexagonal boron nitride. Phys. Rev. Materials 2019, 3, 014004.

[36] Bayer, B. C.; Caneva, S.; Pennycook, T. J.; Kotakoski, J.; Mangler, C.; Hofmann, S.; Meyer, J. C. Introducing overlapping grain boundaries

in chemical vapor deposited hexagonal boron nitride monolayer films. ACS Nano 2017, 11, 4521–4527.

[37] Valvin, P.; Pelini, T.; Cassabois, G.; Zobelli, A.; Li, J.; Edgar, J. H.; Gil, B. Deep Ultraviolet Hyperspectral Cryomicroscopy in Boron Nitride: Photoluminescence in Crystals with an Ultra-Low Defect Density. AIP Advances 2020, 10, 075025.

[38] Rousseau, A.; Moret, M.; Valvin, P.; Desrat, W.; Li, J.; Janzen, E.; Xue, L.; Edgar, J. H.; Cassabois, G.; Gil, B. Determination of the Optical Bandgap of the Bernal and Rhombohedral Boron Nitride Polymorphs. Phys. Rev. Materials 2021, 5, 064602.

[39] Moret, M.; Rousseau, A.; Valvin, P.; Sharma, S.; Souqui, L.; Pedersen, H.; Högberg, H.; Cassabois, G.; Li, J.; Edgar, J. H.; Gil, B. Rhombohedral and Turbostratic Boron Nitride: X-Ray Diffraction and Photoluminescence Signatures. Appl. Phys. Lett. 2021, 119, 262102.

[40] Rousseau, A.; Valvin, P.; Desrat, W.; Xue, L.; Li, J.; Edgar, J. H.; Cassabois, G.; Gil, B. Bernal Boron Nitride Crystals Identified by Deep-Ultraviolet Cryomicroscopy. ACS Nano 2022, 16, 2756.

[41] Pelini, T.; Elias, C.; Page, R.; Xue, L.; Liu, S.; Li, J.; Edgar, J. H.; Dréau, A.; Jacques, V.; Valvin, P.; Gil, B.; Cassabois, G. Shallow and Deep Levels in Carbon-Doped Hexagonal Boron Nitride Crystals. Phys. Rev. Materials 2019, 3, 094001.

[42] Rousseau, A.; Valvin, P.; Elias, C.; Xue, L.; Li, J.; Edgar, J. H.; Gil, B.; Cassabois, G. Stacking-Dependent Deep Level Emission in Boron Nitride. Phys. Rev. Materials 2022, 6, 094009.

[43] Gil, B.; Desrat, W.; Rousseau, A.; Elias, C.; Valvin, P.; Moret, M.; Li, J. H.; Janzen, E.; Edgar, J. H.; Cassabois, G. Polytypes of sp2-Bonded Boron Nitride. Crystals 2022, 12(6), 782.

## Methods

**Preparation of Fe-Ni (111) alloy films**

The as-received C-plane sapphire substrates were first ultrasonically cleaned with deionized water, ethanol, deionized water for 10 minutes in sequence. Then sapphire substrates were placed in a magnetron sputtering chamber for Ni deposition with a thickness of 300 nm under an argon atmosphere. The sputtered sapphires were pre-annealed at 1050 °C under a mixed-gas flow ($Ar/H_2$ = 9:1, flow rate = 500 sccm) for 10 minutes to form single crystal Ni (111), and then a Fe film was thermally evaporated onto the Ni (111) surface (Fig. S2 (a)). The Fe content in Fe-Ni alloy was controlled by the Fe source weight (Fig. S5). As the Fe content increases, the XRD peak corresponding to the Fe-Ni (111), which is located around $2\theta$=44 degrees, gradually shifts to the lower angle (Fig. S3), telling an interplanar spacing *d* gradually increases. This is consistent with the fact that Fe doping induces a lattice expansion of Ni.

**Growth of multilayer BN films by chemical vapor deposition**

The multilayer BN films were grown by CVD method using a single zone tube furnace (Fig. S21). The Fe-Ni (111) substrate was placed at the heating zone, and ammonia borane was located upstream as a precursor. The temperature of ammonia borane precursor was controlled by a heating tape. The furnace was first pumped to a pressure of ~ 5 Pa, and then filled with Ar/$H_2$, which were repeated three times. During the growth process, the furnace was first heated to 1000 °C (15 °C/min), then slowly raised to 1050 °C (5 °C/min), and held for about 10 minutes for annealing. After that the precursor was heated to about 120 °C to start the growth under a mixed-gas flow (Ar/$H_2$ = 9:1, flow rate = 500sccm) for 10 minutes. Once growth was complete, the furnace was quickly cooled down to room temperature (Fig. S21 (a)).

**Transfer of multilayer BN films**

The synthesized multilayer BN films were transferred onto $SiO_2$/Si substrates or holey-carbon TEM grids through the polymethyl-methacrylate (PMMA) - mediated transfer method (Fig. S22). Firstly, spin-coating of PMMA (5% PMMA anisole solution) on the surface of sample, and heating at 120 °C for 6 minutes to evaporate the solvent were followed by etching the Fe-Ni alloy in a $FeCl_3$ solution until the PMMA/BN floats. After that, PMMA/BN was picked up into dilute hydrochloric acid to remove $Fe(OH)_3$ colloid, then washed with deionized water for 2-3 times, and picked up with the target substrate ($Si/SiO_2$(90 nm), holey carbon on copper mesh), heated at 120 °C for 10 minutes to dry. Finally, it was soaked in acetone to dissolve the PMMA, and washed with isopropanol.

**Observation of the BN back side**

The synthesized multilayer BN films were transferred upside down onto a glass slide (Fig. S23) for AFM to reveal the structure of multilayer BN. Firstly, spin-coating of PMMA (5% PMMA anisole solution) on the surface of sample, and heating at 120 °C for 6 minutes to evaporate the solvent were followed by reversing the sample and sticking it on a glass slide with UV glue. Afterwards it was put into a $FeCl_3$ solution to etch the Fe-Ni alloy for enough time. Finally, the BN/PMMA/glass slide was picked up into dilute hydrochloric acid to remove the $Fe(OH)_3$ colloid, then washed with deionized water for 2-3 times.

**Characterizations of BN**

Optical images were obtained using an Olympus BH-2 microscope with magnifications of 50X, 200X and 500X. Raman spectra were obtained with a Horiba HR-800 system with a laser excitation wavelength of 532 nm under ~ 45 μW power. XRD data was collected on a Rigaku Smartlab3 using a copper target and the power was 3 KW. SEM images were obtained using a Hitachi S-4500e SEM with an accelerating voltage of 1 KV. AFM images were acquired using a Veeco Dimension 3100 system in tapping mode. TEM observations and SAED were performed using FEI Titan 80-300 with accelerating voltage of 80 - 300 kV. The cross-section TEM sample was fabricated by a Thermo Fisher Helios 5 CX FIB-SEM. First use the electron beam to deposit 0.2 μm Pt, and then use the ion beam to deposit 2 μm Pt, and Pt is used as a protective layer. After the sample was extracted and placed on the lift-out grid, it was thinned with 8kV and 5kV voltage in turn, and finally cleaned with 2kV. The cutting direction is along the BN arm-chair direction (assuming a zig-zag edge of the BN triangle domains), enabling the TEM observation along the zig-zag direction. High-resolution transmission electron microscopy (HRTEM) images were processed by Digital Micrograph software. In order to enhance the contrast, we first select an area for the Fast Fourier transform (FFT), then select the appropriate mask and perform the inverse Fast Fourier transform (IFFT), and finally superimpose with the original image. LEED patterns were obtained by a LEED-Auger spectrometers BDL800IR system in ultrahigh vacuum with a base pressure below 3 × 10$^{-7}$ Pa. The beam spot size is ~500 microns.

The conduction characteristics were performed by cAFM with an Asylum ORCA cantilever holder with a gain of 1 × 10$^{-9}$ V · A$^{-1}$ by using a Pt/Ir coated cantilever probe.

The XPS spectra were obtained with a Thermo Fisher Scientific K-Alpha, using Al target as the excitation source. The spot size for measurement is ~ 400 microns. For the processing of XPS data, we calibrated the binding energy with C1s (284.8 eV), Then N1s and B1s were found to be well aligned at 398.08 eV and 190.53 eV, respectively. The Ni2p spectra displays two spin–orbit doublets located at 852.4 eV (Ni2p$_{3/2}$) and 869.7 eV (Ni2p$_{1/2}$) originate from metal state nickel. A weak peak at 855.9 eV is observed in the 31% Fe sample, corresponds to Ni$^{2+}$/Ni$^{3+}$, indicating the existence of the metal oxide. No such peak is observed in the 51% Fe sample. We thus contribute the observation of metal oxide to a smaller BN coverage, which is well fitting to the AFM and LEED results. The Fe2p spectra also display two spin–orbit doublets located at 706.6 eV (Fe2p$_{3/2}$) and 719.8 eV (Fe2p$_{1/2}$), however without evident oxidation states. As a matter a fact, the Fe2p spectra show relatively weak peaks but with evident satellite shoulder at high energy side, which makes the observation of metal oxide a little difficult. By distracting the XPS baseline (Shirley back-ground) and normalizing by the Fe2p$_{3/2}$ peak, we find the 31% Fe sample shows significantly higher intensity around the Fe2p$_{3/2}$ satellite (Fig. S10 (g)), which may be due to the superposition of the satellite peak and the iron oxide peak (binding energy of Fe$_2$O$_3$ peak is ~ 710.8 eV). The automatic overlapping of the Fe2p$_{1/2}$ peak validates the viability of this data manipulation (Fig. S10 (g)). We thus would like to conclude a high tendency that the 31% Fe sample is more evidently oxidized. Such a conclusion also supports the change of BN coverage as discussed in AFM and LEED.

**Spectroscopy in the deep ultraviolet**

Photoluminescence (PL) spectroscopy and Second Harmonic Generation (SHG) were performed with the setup described in ref. 37. Briefly, PL excitation is provided by the fourth harmonic of a continuous-wave (cw) mode-locked Ti:Sa oscillator and it is tunable from 193 nm to 205 nm with trains of 140-fs pulses at 80 MHz repetition rate. SHG is excited by the second harmonic of the Ti:Sa oscillator. The injection path uses a series of metallic and dichroic mirrors coated for these spectral UV ranges. The exciting laser beam is focused by a Schwarzschild objective located inside the closed-cycle cryostat equipped with CaF$_2$ optical windows. The objective numerical aperture is 0.5, resulting in a diffraction-limited laser spot size of 200 nm in PL spectroscopy. The sample is mounted on a stack of piezoelectric steppers and scanners, which is cooled down to 6 K under ultrahigh vacuum ($10^{-11}$ bar). The emission is collected by means of an achromatic optical system comprising a pinhole for confocal filtering. It is then dispersed in a Czerny−Turner spectrometer with 500 mm focal length and a 1200 grooves/mm ruled grating and finally detected by a back-illuminated CCD camera with 13.5 μm pixel size. The corresponding spectral resolution is 0.17 nm (3.3 meV) at 200 nm. The entire setup is controlled by homemade Python modules inserted inside the QUDI software suite33 to create hyperspectral images of the PL signal recorded at each point of a scanned area.

**Supporting information**

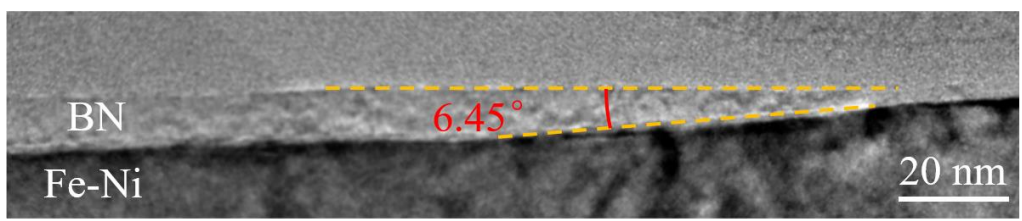

*Fig. S1 | Cross-sectional transmission electron microscope image of the boron nitride (BN) film grown on Fe-Ni substrate. The angle indicated is consistent with the AFM results in Figure 1.*

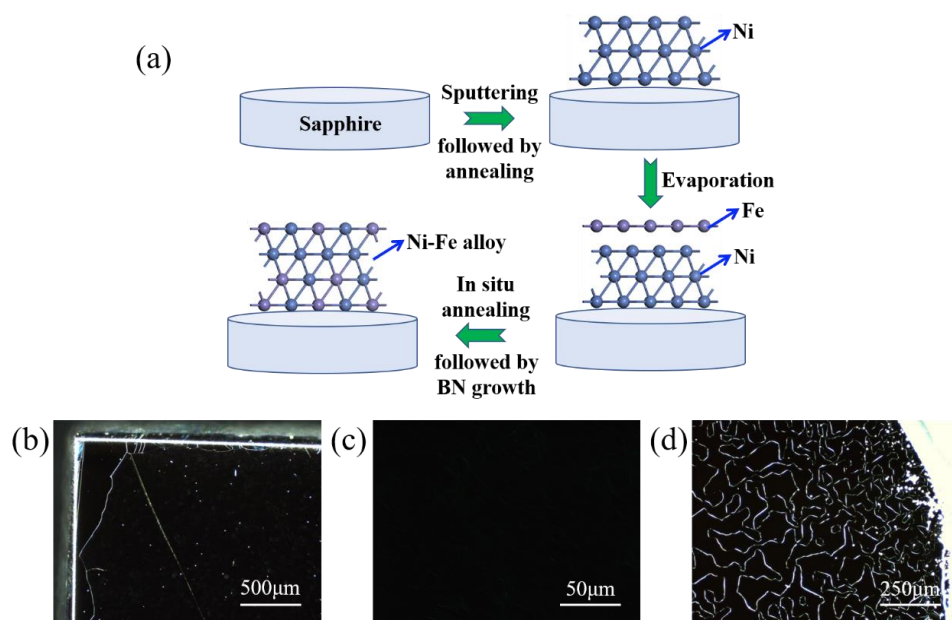

*Fig. S2 | Preparation of the Fe-Ni (111) substrate. (a) Schematic diagram of the preparation process of Fe-Ni alloy films. (b) (c) The optical dark field images of the single-crystal Ni (111) with few (b) or without (c) twin boundaries before the growth of boron nitride. (d) The optical dark field images of the Ni (111) with twin boundaries before the growth of boron nitride. Note the bright lines are twin boundary among Ni (111) anti-domains.*

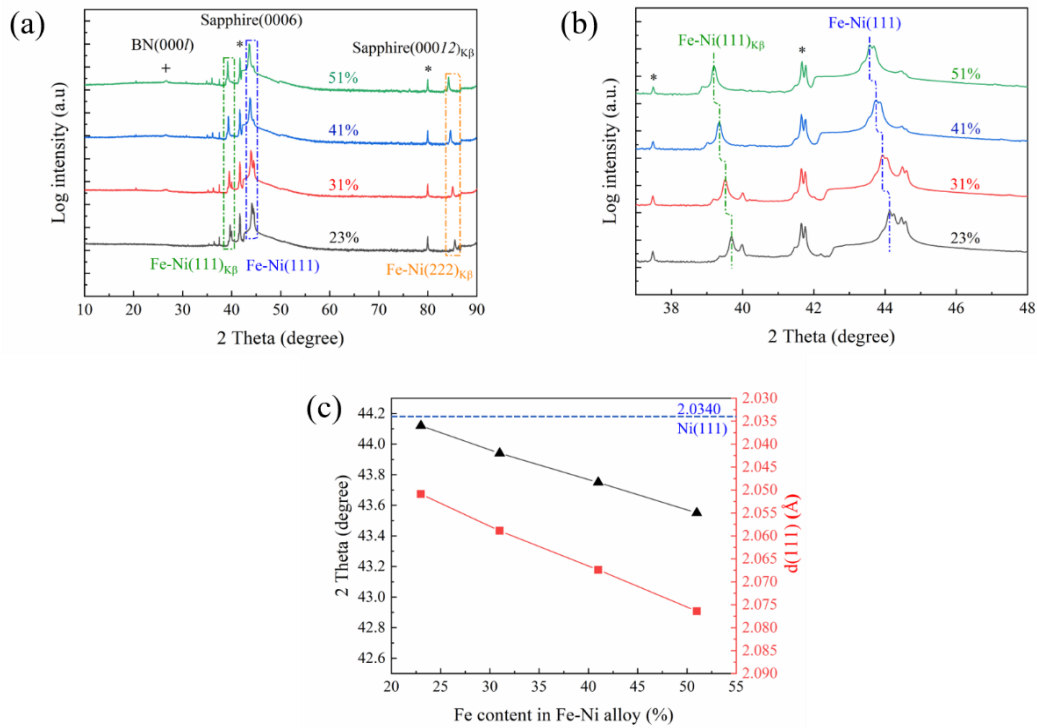

*Fig. S3 | The X-ray diffraction results of the obtained Fe-Ni alloy film. (a) The X-ray diffraction pattern in the 2θ scan. The peaks related to sapphire are marked with '*', the peak of boron nitride is marked with '+', and the peaks of Fe-Ni alloy film are framed with dotted lines. Fe content is noted on the data. (b) Enlarged view of the peaks of the Fe-Ni alloy film. (c) Peak position and change of d (111) with Fe content. The d (111) of pure Ni, 2.0340 Å, is marked with a blue dashed line. Note the d (111) axis increase from the top to the bottom.*

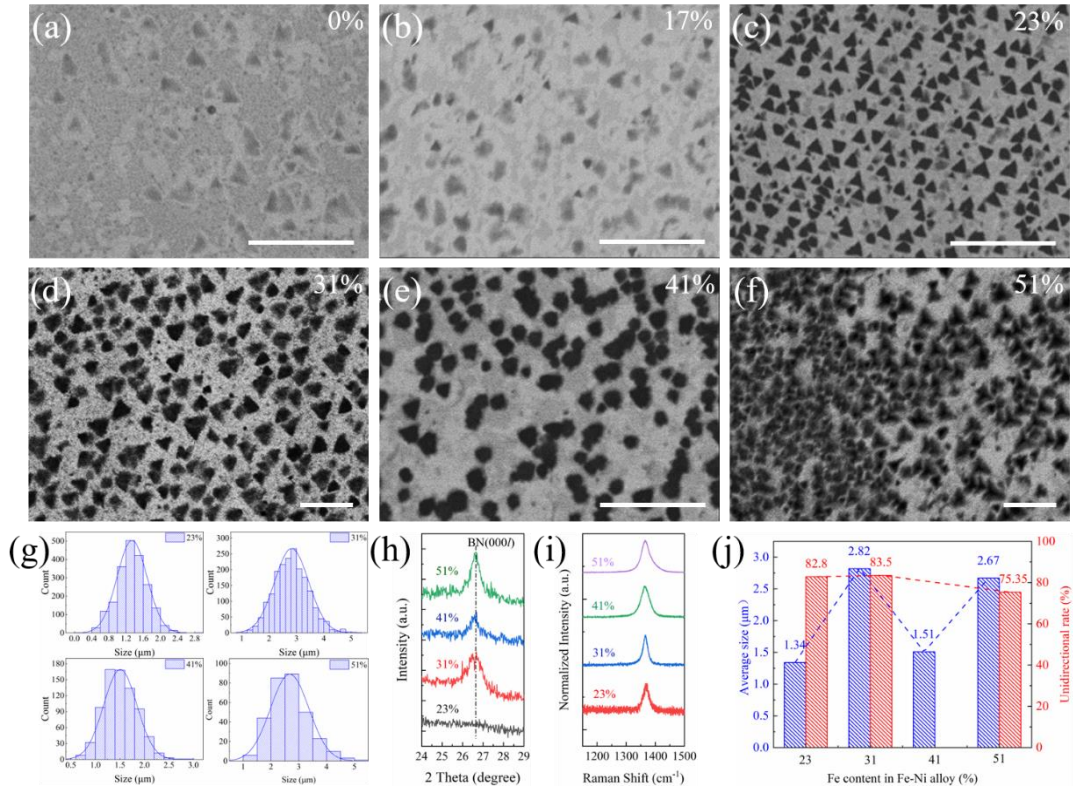

*Fig. S4 | The characteristics of boron nitride (BN) grown on alloy films with different Fe content. (a) 0% Fe. (b) 17 Fe%. (c) 23%, which is a repeat of Fig. 1 (g). (d) 31%, which is a repeat of Fig. 1 (h). (e) 41% Fe%. (f) 51 Fe%. The scale bar is 10 microns. (g) The statistical histogram of the size of BN domains. (h) The X-ray diffraction pattern in the θ-2θ scan. (i) The typical Raman spectrum of BN on Fe-Ni (111). (j) Average size and unidirectional rate of BN domains with different Fe content.*

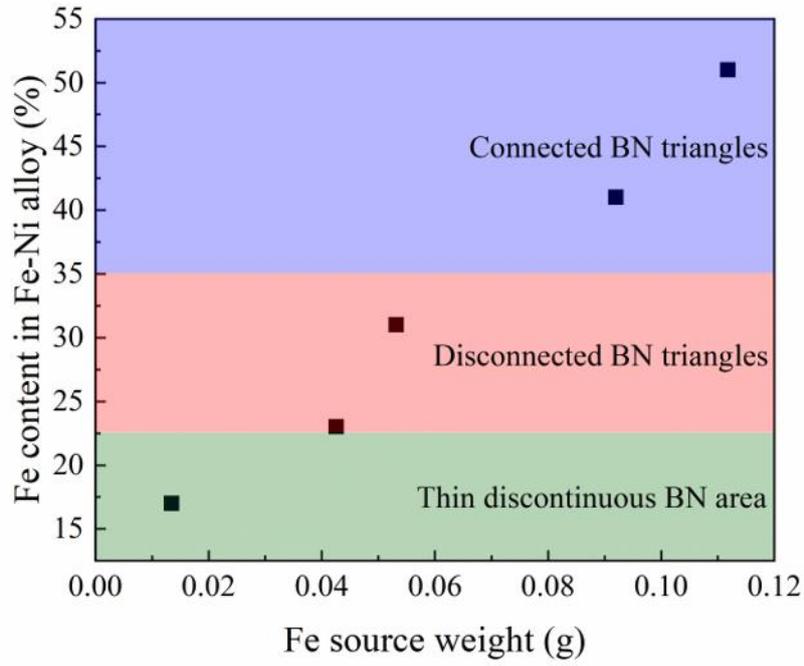

*Fig. S5 | The plot of Fe content in Fe-Ni alloy versus the evaporated Fe source weight. The Fe content was given by energy dispersive spectroscopy (EDS).*

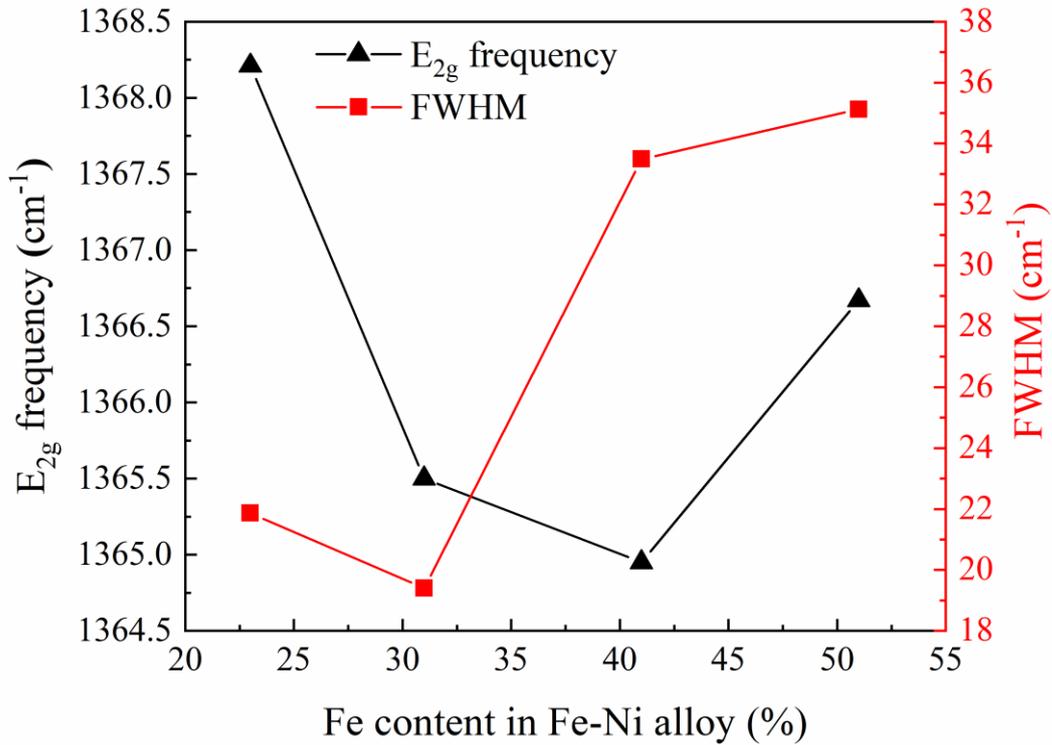

*Fig. S6 | Line graph of $E_{2g}$ peak position and FWHM versus Fe content.*

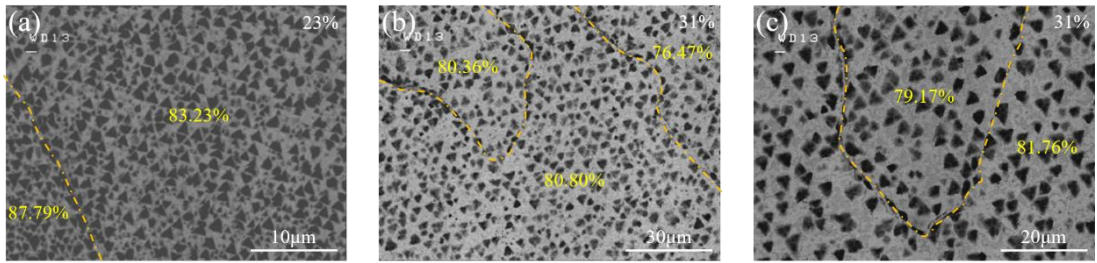

*Fig. S7 | Statistics of uni-orientation on different grains of the Fe-Ni alloy films. (a) The sample with 23% Fe. (b), (c) The samples with 31% Fe separate runs of growth, indicating the good reproduction. The dashed lines indicate the grain boundaries in Fe-Ni (111) thin film [13].*

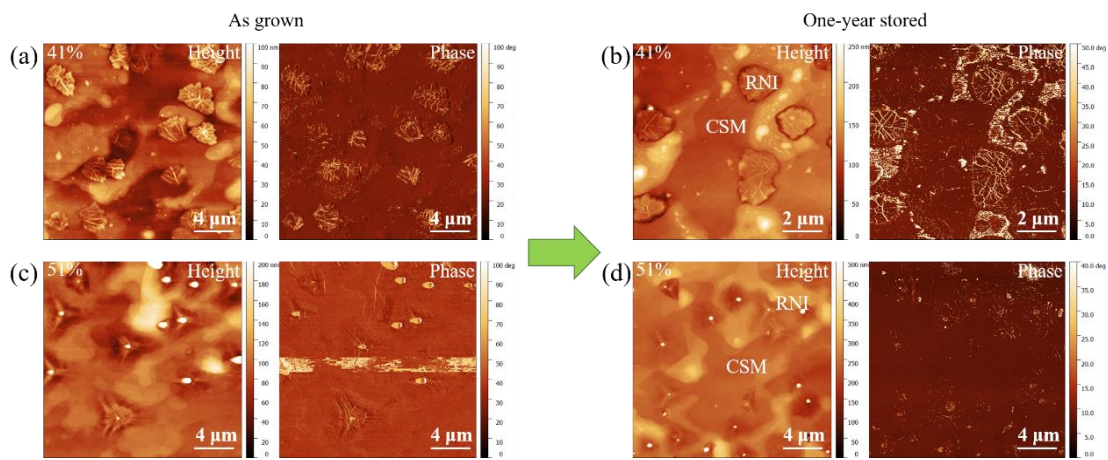

*Fig. S8 | The atomic force microscope images of as-deposited samples and that stored for 1 year with Fe content of (a) (b) 41% and (c) (d) 51%.*

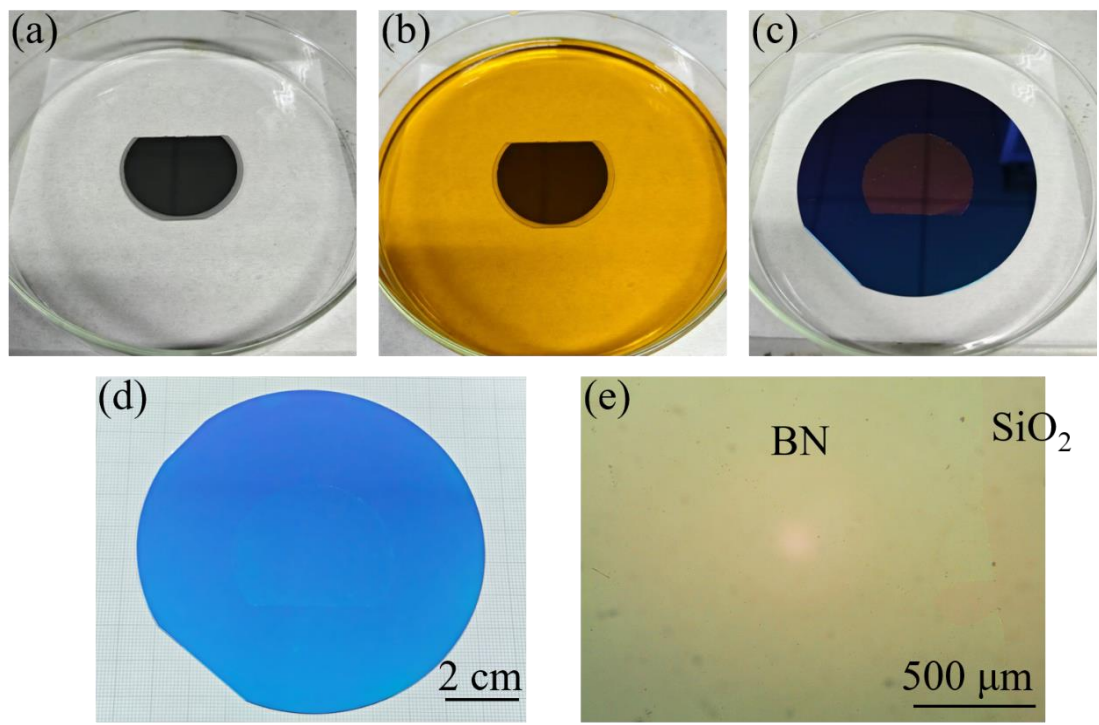

*Fig. S9 | Wafer-scale boron nitride (BN) multilayers transfer process. (a) the 51% Fe sample with BN and PMMA films. The wafer size is two inch, and the edge was cut in order to put the wafer into the furnace tube. (b) The sample was immersed in the $FeCl_3$ solution to etch the Fe-Ni. (c) The BN /PMMA films were transferred onto a four-inch SiO2/Si (300 nm*

$SiO_2$) substrate, and dried at a low temperature. (d) The PMMA was removed by acetone for one hour, and the optical photograph shows the intact and homogeneous BN films. (e) Optical microscope image of the transferred continuous BN films.

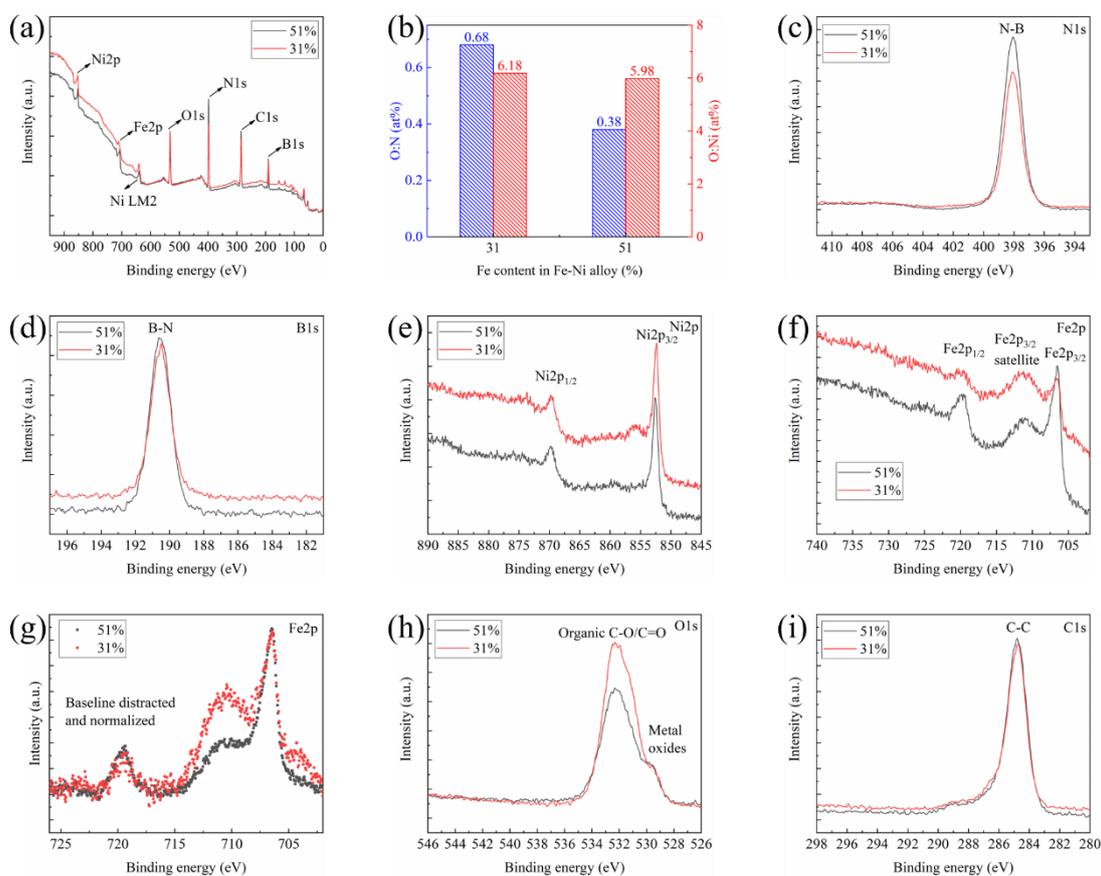

Fig. S10 | *The X-ray photoelectron spectroscopy (XPS) measurement of samples with different Fe contents. (a) The XPS survey of the samples of 31% Fe and 51% Fe. (b) The atomic ratios of O to N and O to Ni were calculated, respectively. With the increase of Fe content, the oxygen content of the sample surface decreased significantly, which was consistent with the AFM results in Fig. 2 and Fig. S8, indicating that*

*boron nitride domains connect to form films and higher coverage inhibits the oxidation of the alloy. (c) The N1s spectra is identified to be a N–B peak at (398.08 eV). (d) The B1s spectra is identified to be a B–N peak at 190.53 eV. (e) The Ni2p spectra display two spin–orbit doublets located at 852.4 eV ($Ni2p_{3/2}$) and 869.7 eV ($Ni2p_{1/2}$), with a weak peak at 855.9 eV of 31% Fe sample, which may correspond to $Ni2+/Ni3+$. (f) The Fe2p spectra display two spin–orbit doublets located at 706.6 eV ($Fe2p_{3/2}$) and 719.8 eV ($Fe2p_{1/2}$). (g) After backline distraction and normalization, the satellite intensity of 31% Fe is significantly higher, which may be due to the superposition of the $Fe2p_{3/2}$ satellite peak and the iron oxide peak, since the binding energy of $Fe_2O_3$ peak is ~710.8 eV. (h) 31% Fe has more organic C-O/C=O. (i) The binding energy was calibrated with C1s (284.8 eV).*

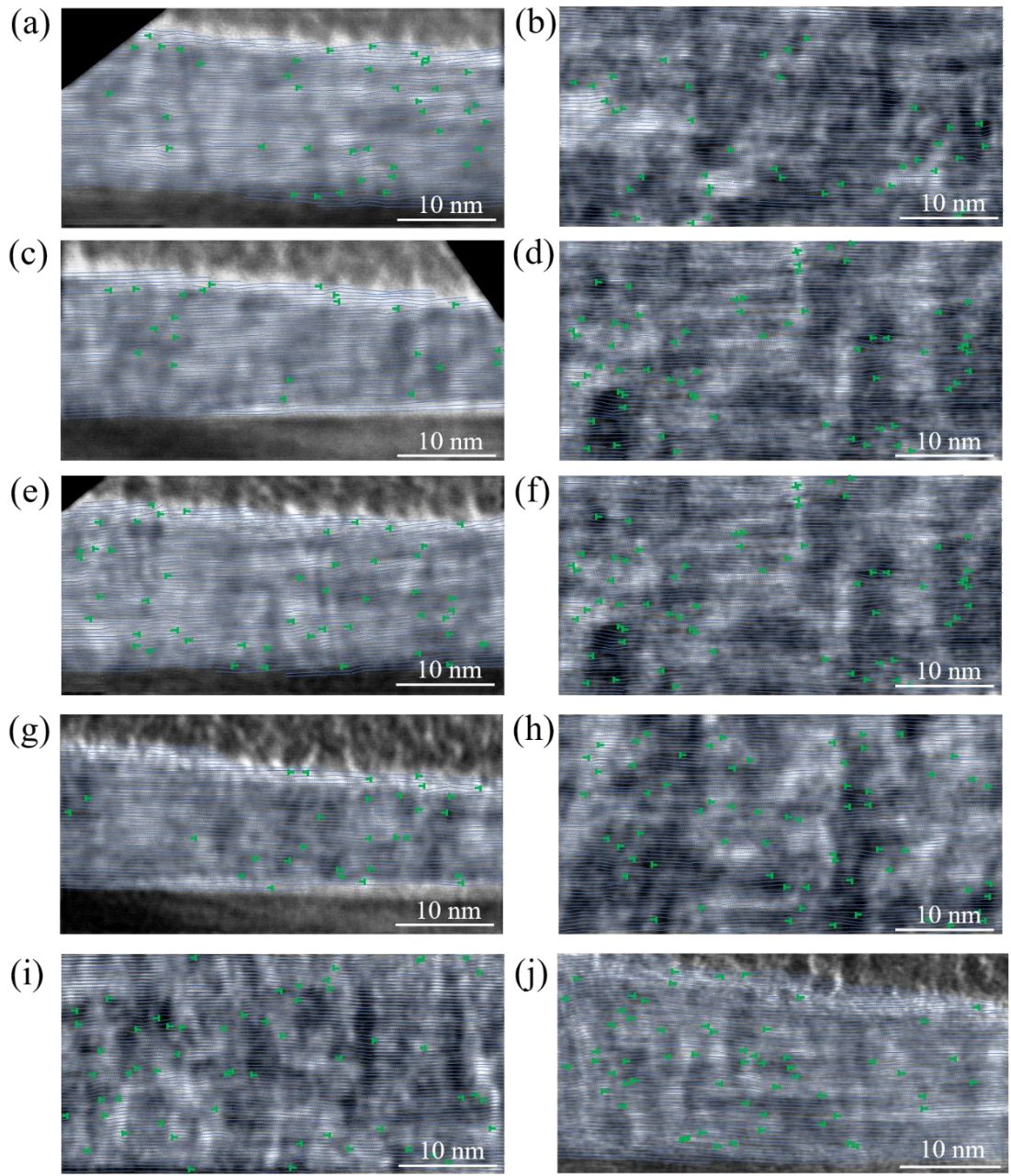

*Fig. S11 | The dislocation analysis of cross-sectional high-resolution transmission electron microscopy (HRTEM) images in randomly chosen individual domain region, in which blue lines (which can be seen clearly when the pictures are enlarged) are guiding the BN layers, and the dislocations (b= [0001]) are marked with the green symbols.*

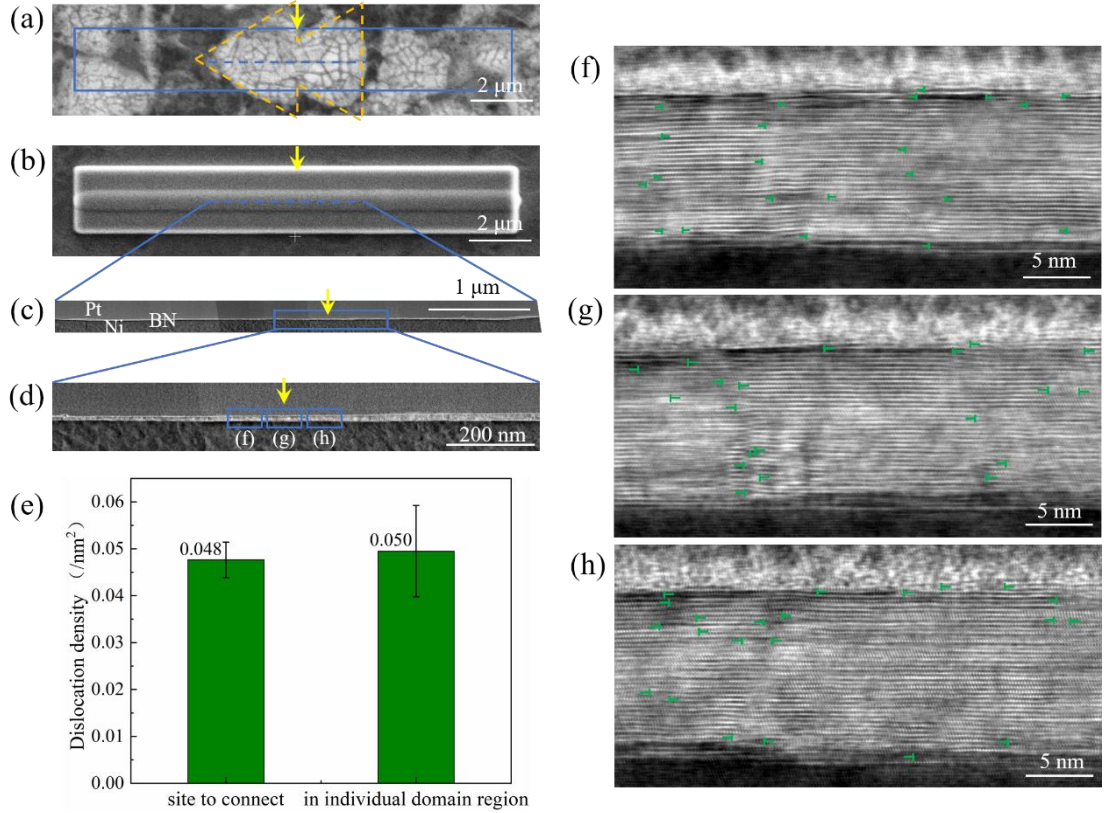

*Fig. S12 | Focused ion beam (FIB) preparation process and statistical analysis of dislocation density of another boron nitride (BN) multilayers sample. (a) Scanning electron microscope (SEM) image of the area chosen for the cross-sectional transmission electron microscopy (TEM) sample, in which the bule line frames the area for Pt deposition, the orange line frames the two triangular BN domains for the cross-sectional observation, and the yellow arrow mark the site to connect of two triangular domains. (b) The SEM image after Pt deposition. (c), (d) Cross-sectional TEM image of the site to connect at increasing magnifications. (e) Statistical histogram of dislocation density (b= [0001]) of the site to connect and inside individual BN domain region. The error bar is the standard deviation. (f) – (h) The cross-sectional high-*

54 / 62

resolution transmission electron microscopy (HRTEM) images of the site to connect. The dislocations (b= [0001]) are marked with the green symbols, and the relative positions are shown in (d).

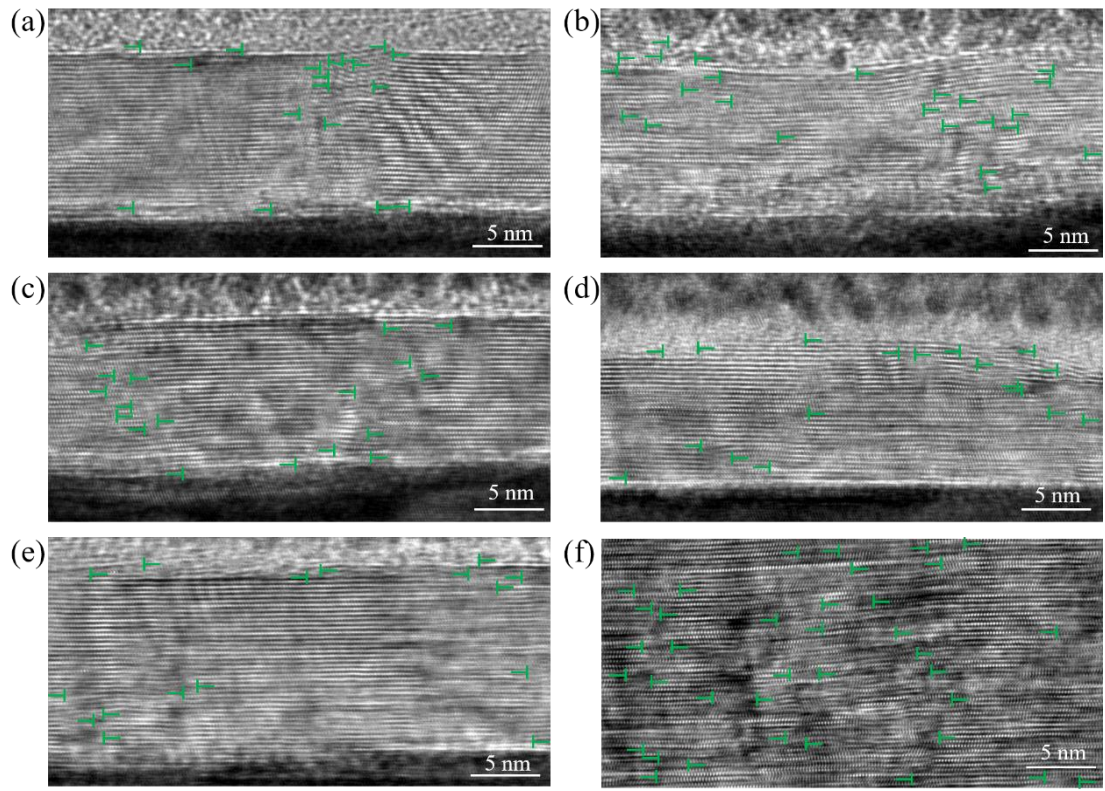

Fig. S13 | The dislocation analysis of cross-sectional high-resolution transmission electron microscopy (HRTEM) images in randomly chosen individual domain region (the sample in Fig. S12), in which the dislocations (b= [0001]) are marked with the green symbols.

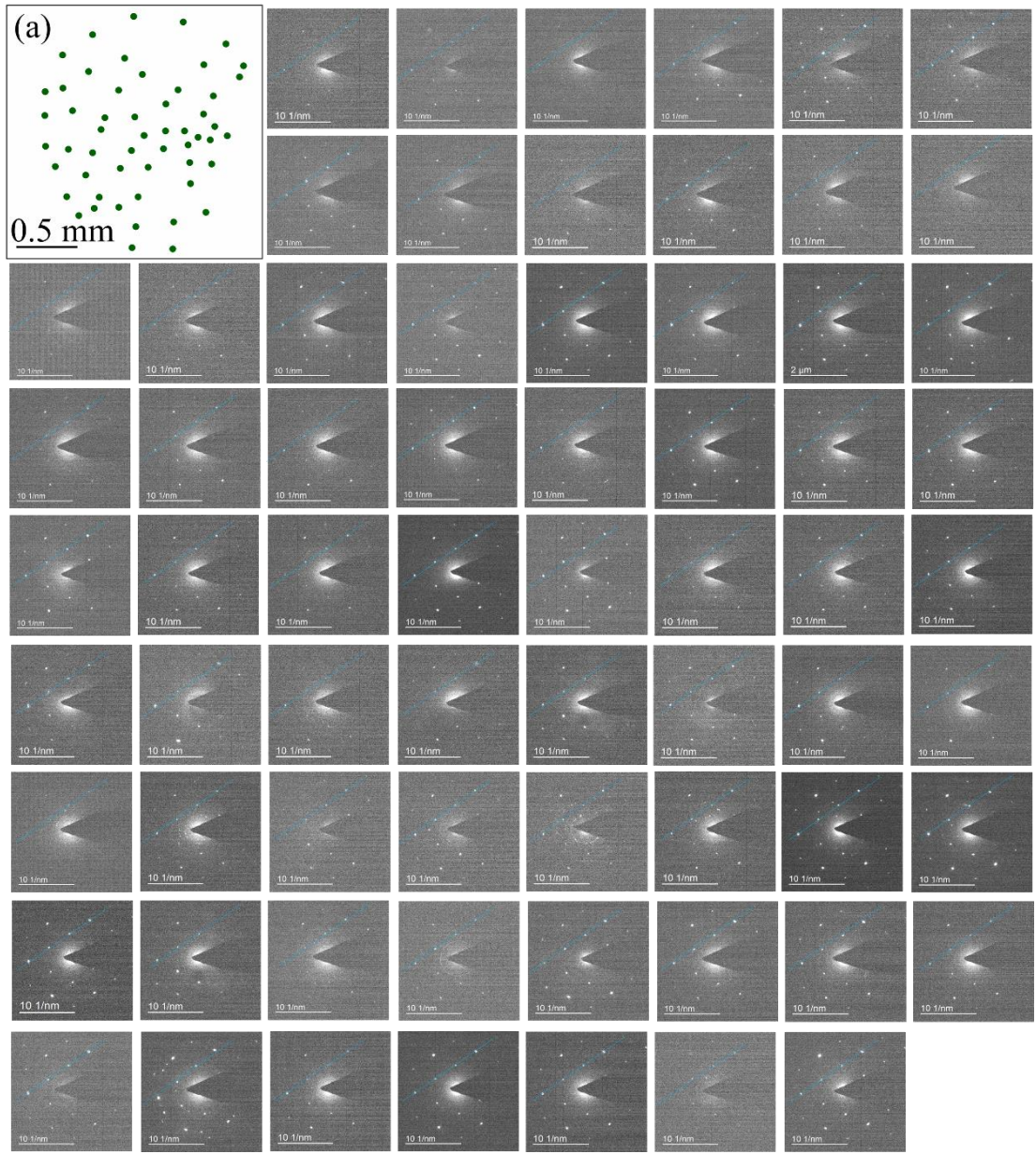

*Fig. S14 | The selected area electron diffraction (SAED) of the synthesized BN films. (a) Global map of the randomly selected points in a large area, 2 mm×2.2 mm. All of the SAED patterns has the same direction (marked with blue lines), proving that the BN orientation is the same over a large area.*

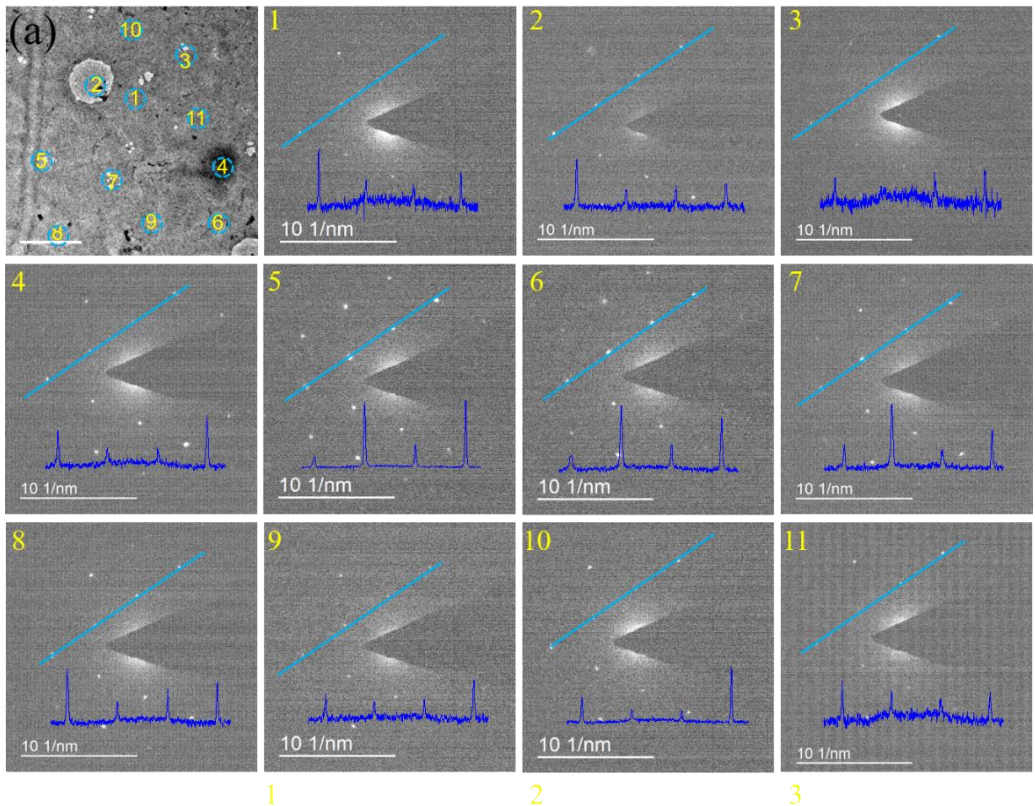

*Fig. S15 | TEM morphology images and selected area electron diffraction patterns of the synthesized boron nitride. (a) A selected flat area of boron nitride films with eleven selected area electron diffraction patterns (1~11).*

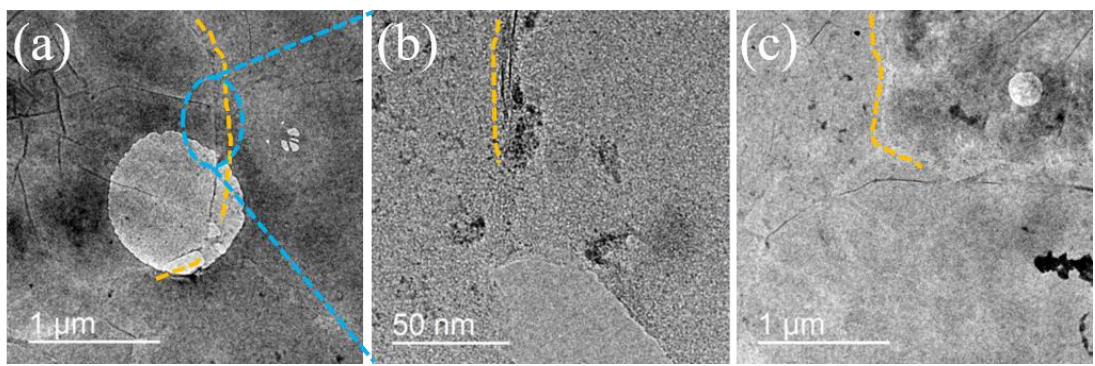

*Fig. S16 | Transmission electron microscope morphology of the boron nitride films. Dashed lines indicate places with contrast variations.*

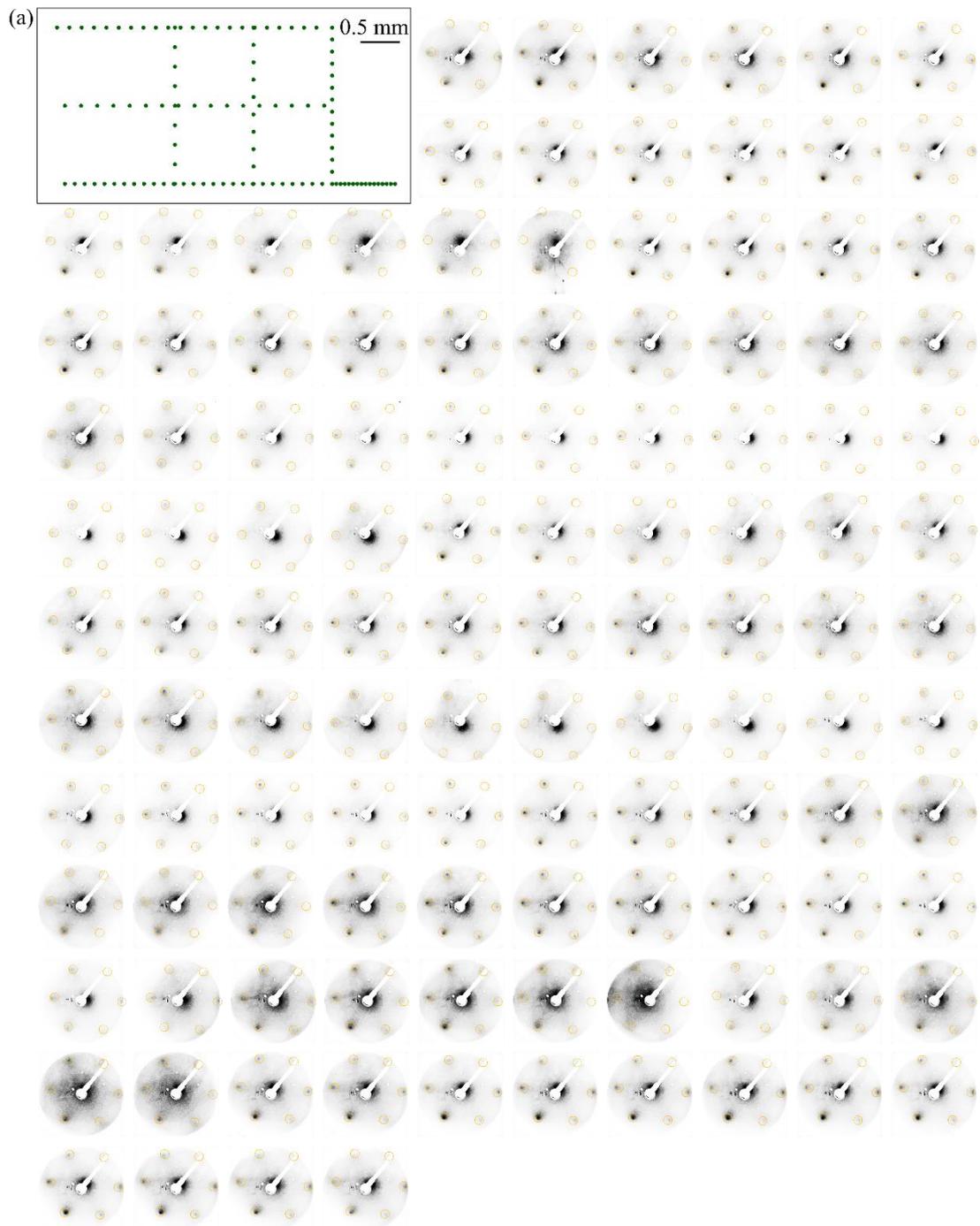

*Fig. S17 | The low-energy electron diffractions (LEED) of 51% Fe sample. LEED patterns were taken at 63eV. (a) Global map of the selected points in a large area, ~2 mm×3 mm. A total of 116 points diffraction orientations are completely consistent, confirming that the*

*crystalline lattice of the BN multilayers is aligned in the same direction.*

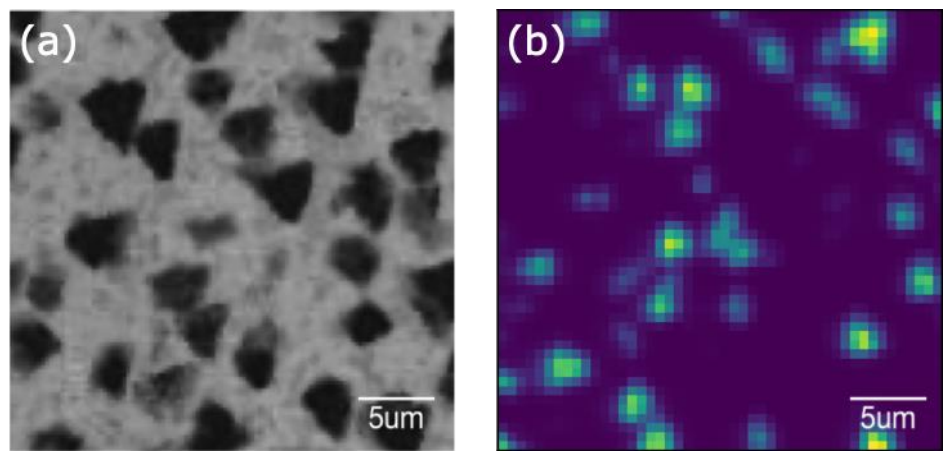

*Fig. S18 | Second harmonic generation (SHG) microscopy in multilayer BN grown on 31% Fe-substrate. (a) typical SEM image. (b) SHG intensity map recorded in a different region, but of the same area as the one in (a). The scalebar in (a), (b) is 5μm.*

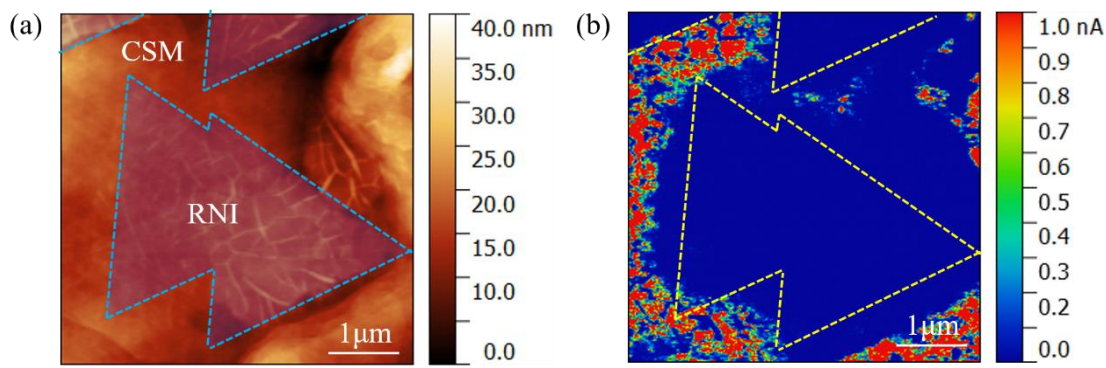

*Fig. S19 | Conductive atomic force microscope (cAFM) measurement of boron nitride (BN) under the bias voltage of 6 V, and the BN film shows small leakage with pico-Ampere level under bias of 6 V. (a) the cAFM height image. (b) The cAFM current mapping. The residue non-connecting island (RNI) are marked with the false-color (blue), and the*

*other regions correspond to the continuously connecting single-crystal multilayer (CSM).*

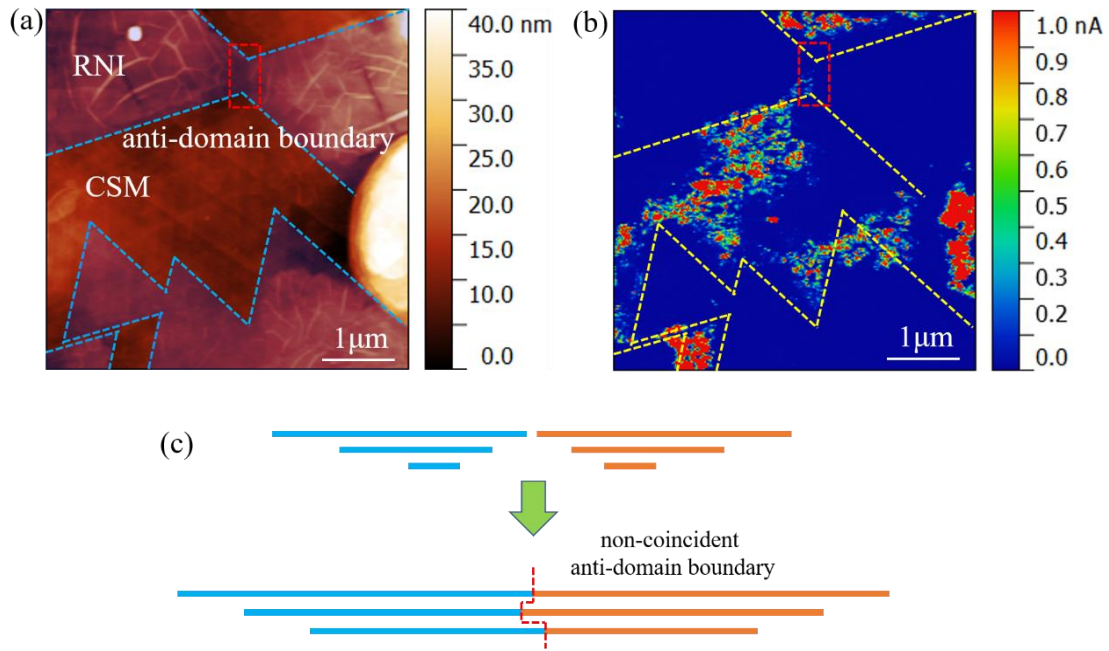

*Fig. S20 | Conductive atomic force microscope (cAFM) measurement of boron nitride under the bias voltage of 6 V. (a) the cAFM height image. (b) The cAFM current mapping. The residue non-connecting island (RNI) are marked with the false-color (blue), and the other regions correspond to the continuously connecting single-crystal multilayer (CSM), and the anti-domain boundary is marked with red dotted lines. (c) The schematic diagram of non-coincident anti-domain boundary, which may also work for effective screening the leaks path.*

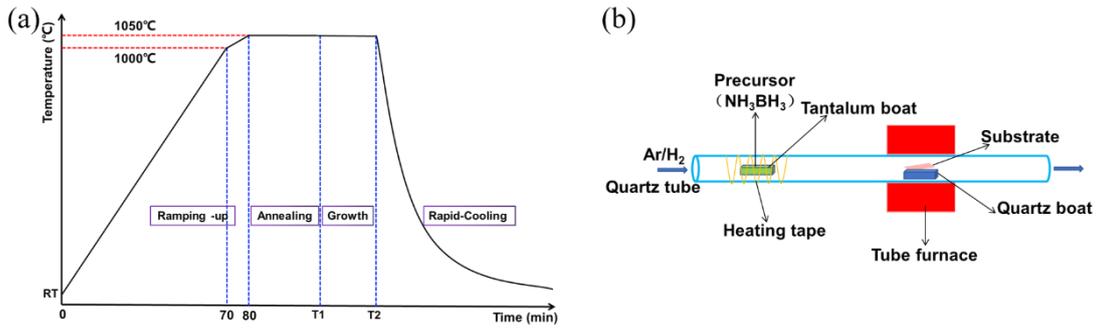

*Fig. S21 | Growth of multilayer boron nitride films by chemical vapor deposition method. (a)Variation of substrate temperature throughout the growth process. (b)Schematic diagram of the growth equipment and conditions.*

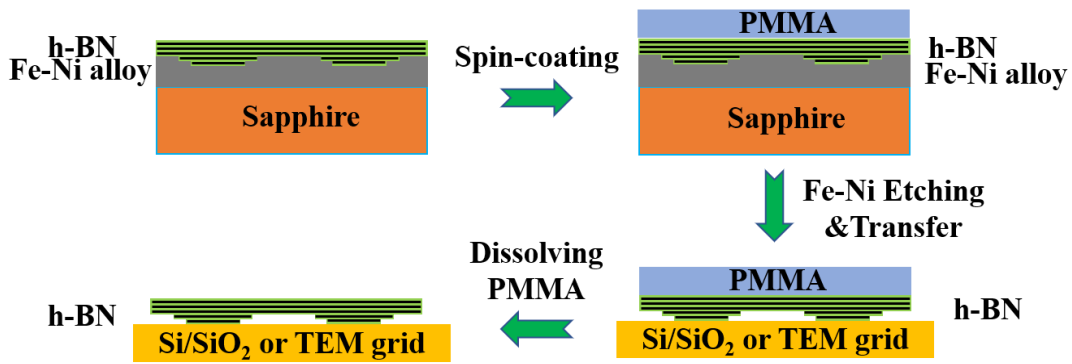

*Fig. S22 | Flow chart of BN transfer to target substrates.*

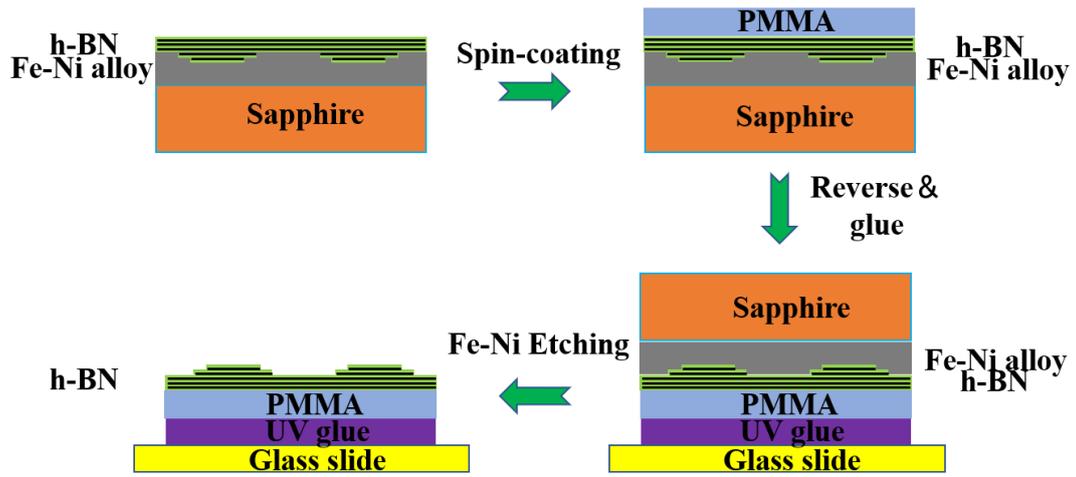

*Fig. S23 | Flow chart of boron nitride transferring for the observation of the BN domain backside structure.*